\documentclass[reqno, 12pt]{article}
\usepackage[T1]{fontenc}
\usepackage[utf8]{inputenc}
\usepackage{tikz}
\usepackage{amssymb}
\usepackage{amsfonts}
\usepackage{amsmath}
\usepackage{scalefnt}
\usepackage{fancybox}
\usepackage{float}
\usepackage{natbib}
\usepackage{theorem}
\usepackage{lscape}
\usepackage{hyperref}
\usepackage{graphicx}
\usepackage{appendix}
\usepackage{caption}
\usepackage{subcaption}
\usepackage{multicol}
\usepackage{multirow}
\usepackage{colortbl}
\usepackage{latexsym}

\newtheorem{theorem}{Theorem}
\newenvironment{theorem*}{\vskip 0.1in \noindent {\bf Theorem \/}}{\vskip 0.1in}

\newtheorem{claim}{Claim}

\newtheorem{definition}{Definition}
\newtheorem{example}{Example}

\newtheorem{proposition}[theorem]{Proposition}
{\theorembodyfont{\upshape} }

\newenvironment{proof}[1][Proof]{\textbf{#1.} }{\ \rule{0.5em}{0.5em}}
\newtheorem{Assumption}{Assumption}

\renewcommand{\cite}{\citet}

\newcommand{\ul}{\underbar{u}}
\newcommand{\uh}{\bar{u}}
\newcommand{\bfP}{\mathbf{P}}
\newcommand{\betaBar}{\Bar{\beta}}
\newcommand{\invPhi}{\Phi^{-1}}

\begin{document}

\title{Identification of Incomplete Preferences}
\author{Arie Beresteanu\thanks{%
Department of Economics, University of Pittsburgh, arie@pitt.edu. } \and %
Luca Rigotti\thanks{%
Department of Economics, University of Pittsburgh, luca@pitt.edu.}}
\date{February, 2025\thanks{%
We thank Miriam Blumenthal for excellent RA work, Pawel Dziewulski for very careful comments on an early draft, Ariel Rubinstein and Richard Van Weelden, as well as audiences at Cornell, Duke, IMSI (Chicago), Penn State, Pitt, SUNY Albany, Tel Aviv, Syracuse, Northwestern.}}
\maketitle

\begin{abstract}

We provide a sharp identification region for discrete choice models where consumers' preferences are not necessarily complete even if only aggregate choice data is available. Behavior is modeled using an upper and a lower utility for each alternative so that non-comparability can arise. The identification region places intuitive bounds on the probability distribution of upper and lower utilities. We show that the existence of an instrumental variable can be used to reject the hypothesis that the preferences of all consumers are complete. We apply our methods to data from the 2018 mid-term elections in Ohio.

\bigskip


\bigskip

\end{abstract}

\thispagestyle{empty}

\setlength{\baselineskip}{.26in}\newpage

\setcounter{page}{1}


\section{Introduction}
\label{introduction}

Since McFadden's 1974 paper, discrete choice models have been a cornerstone of applied economics analysis. These models, like much of economics, assume that individuals' preferences are complete: individuals can always rank all alternatives. However, there are important reasons why individuals' preference may not be complete; for example, one may have to choose without having enough information about each of the alternatives, or one may have to choose without having enough information about one's own preferences. Assuming this issue away in identification problems is convenient because completeness induces a definite choice, and therefore makes identification of rational behavior easier.\footnote{Indifference implies alternatives are ranked as equal, and is typically deemed a zero probability occurrence.} In this paper we tackle identification when individuals' preferences are allowed to be incomplete even if only aggregate behavior is observable (or, equivalently, individuals make only one observable choice). This is a classic discrete choice setting modified so that completeness does not necessarily hold. The objective is to identify properties of the probability distribution of preferences across heterogeneous individuals using either aggregated market level or individual choice data.

Our main results contradict two hypotheses one could make about the relationship between data and theory when preferences are not complete. The first is that without completeness anything can happen; data cannot tell us anything about the underlying preferences unless one makes assumptions about how choices might result when alternatives are not comparable. The second is that completeness cannot be ruled out by observable behavior. Pairing insights from econometric theory and decision theory, we show that parameters of the probability distribution of preferences across individuals are partially identified, and thus data restricts the possible underlying preferences even if nothing is assumed about how choices among non comparable alternatives are made. We then illustrate how data could show that some of the individuals in the population must have incomplete preferences, thus dispelling the second hypothesis. We also show that if preferences are not complete, a policy aimed at improving the frequency with which a particular alternative is chosen may actually have the opposite effect –a result that can only happen if preferences are not complete.

When preferences are not complete an individual may not be able to rank some pairs of alternatives. Therefore, observers cannot infer that a chosen alternative must have been at least as good as those not chosen. They can only infer that the alternatives not chosen could not have been strictly preferred to the chosen one. When using data to learn about the distribution of preferences in the population, allowing for incompleteness poses what may seem like an insurmountable problem. Because the theory says nothing about how choices between incomparable alternatives are made, data cannot help distinguish choices made because of a definite preference from choices made randomly when alternatives are not comparable. In standard discrete choice models this source of randomness is ruled out by assuming everyone's preferences are complete. One may suspect that when completeness is not imposed data cannot say anything about preferences. To the contrary, we show that also in this case data can be used to learn about the distribution of preference parameters. Obviously, identification has weaker properties than in the case of complete preferences.

We consider a setting in which the analyst has data on the choices of a population of heterogeneous rational agents who must choose one alternative from a given feasible set. With complete preferences, each alternative is associated with a utility value that is known to the decision maker but not known to the analyst. The analyst, on the other hand, knows from data the frequency with which each alternative is chosen. These observed choices are used to make inferences about the probability with which one option is preferred to the others in the population. Since completeness implies each alternative is associated with a utility value, rationality implies that the proportion of individuals who chose one alternative must be equal to the proportion of individuals who assign higher utility to that alternative. 

We model incompleteness by allowing for the possibility that each alternative is associated with many utility values. An example of this framework is given by the \emph{interval order} of \cite{Fishburn1970book}, where each alternative is associated with an interval of utilities; other possible examples, in stochastic environments, are the multi-utility models of \cite{Aumann62}, \cite{Ok02}, and \cite{DubraMaccheroniOk04}, or the multi-probability model of \cite{Bewley02}; more recent examples combine Fishburn's and Bewley's models like \cite{Echenique-Pomatto-Vinson} and \cite{Miyashita-Nakamura}. We focus on the simplest of these settings, where alternatives are compared looking at the extremes of their utility intervals: their upper and lower utility. An alternative is better if its lower utility is larger than the upper utility of the other, worse if its upper utility is smaller the the lower utility of the other, and not comparable when neither of these conditions is satisfied. 

Rationality implies that when two utility intervals are disjoint, the corresponding alternatives can be ranked and choice follows this ranking; when two intervals overlap, the corresponding alternatives are not comparable and choice is indeterminate. With more than two alternatives, rationality implies an alternative can be chosen as long as its upper utility is larger than the lower utility of all other alternatives. In other words, rationality means that decision makers choose an alternative that is not strictly dominated by any other in the feasible set. The choice set is thus the set of non-dominated alternatives; this is a singleton when preferences are complete, while it can contain many elements when they are not. Without completeness, the link between observed choice data and preferences is therefore weakened. 

We assume there is a population of decision makers with potentially incomplete preferences from which a random sample is drawn. Although we relax completeness, we make no behavioral assumptions other than rationality. The choice set, the set of non-dominated alternatives, is a subset of the feasible set that is not necessarily a singleton. Because the sample is random and choice between non comparable alternatives is also random, the choice set is a random set and rationality implies that the observed choice is a random selection from this random set. Using data on choice probabilities, the parameters one would like to identify are the probabilities that the set of non-dominated alternatives is equal to each of the possible subsets of the feasible set. These parameters describe the distribution of (incomplete) preferences in the population.

Our first main result shows that although point identification is not possible in this setting, partial identification is. In particular, we characterize the identification region for the parameters of interest using a finite number of inequalities relating these parameters to the observed choice probabilities. When the feasible set contains only two options, the identification region for the probability an alternative is the preferred one is easy to describe with an interval. At one extreme, all individuals who choose an option do so because it is not comparable to the other, and thus the lower bound of the interval is zero. At the other extreme, rationality implies that the upper bound of the probability an alternative is the preferred one cannot exceed the fraction of consumers who chooses it. Since all individuals who prefer an alternative choose it (and some individuals might choose it even though they cannot compare it the other), the probability that a randomly selected individual prefers this alternative cannot exceed the observed fraction of individuals who chooses it. When the feasible set contains more than two options, the sharp identification region is harder to describe because it depends on more than these two inequalities as it can be a strict subset of the interval we just described.

Without additional assumptions one cannot rule out the possibility that non-comparability never occurs because all individuals who choose an alternative do so because it is ranked better than all others. This implies incompleteness is either absent or it is present but irrelevant to behavior. Our second main result uses the notion of an instrumental variable to rule out this possibility. We define an instrument as a random variable that is independent of preferences but correlated with choice; in particular, the fraction of individuals who choose an alternative changes for each realization of the instrument while preference characteristics do not. If such an instrumental variable exists there must be some individuals who have incomplete preferences. Intuitively, if the fraction of individuals who choose a certain alternative changes with the realization of the instrument, it cannot be that all individuals are always able to rank all alternatives. Existence of an instrument thus establishes that there must be some incomparable alternatives, making the identification region smaller.

We analyze the impact of a policy intervention aimed at increasing the chances a particular alternative is chosen by increasing the utility individuals give to that alternative. Examples of such a policy could be campaign advertising, or other forms of information provision, that target a particular alternative. When preferences are complete, one can easily show that increasing the utility of an alternative necessarily increases the frequency with which that alternative is chosen. Without completeness, this is no longer the case. Intuitively, unless the intervention is particularly strong, because of incompleteness the behavior of some individuals cannot be predicted even after the policy is enacted. There could be enough individuals that change behavior in favor of the non-targeted alternative to make the policy completely ineffective. Therefore, if one observes aggregate behavior go in the opposite direction than the one the intervention tried to achieve one cannot necessarily conclude the policy was ill-designed.

We study several extensions of the basic model that can also make the identification region smaller. First, we examine a model in which a known fraction of individuals pays attention only to a subset of the feasible set. Second, we show how minimizing ex-post regret could bring point identification. Third, in an extension that is particularly relevant for our application, we consider the situation in which the choices of a known fraction of the population are not observed.\footnote{In our application, voting, a known fraction of individuals go to the polls but abstain in some of the races on the ballot.} In this case, in addition to partial identification resulting from incomplete preferences, researchers can face non response or unobserved choices. We show that the identification region can then be written as a convex combination of an identification region resulting from incomplete preferences and an identification region resulting from non response. Finally, while our results are cast in a world without uncertainty, we show that one can allow for ambiguity by extending our framework to study identification of the preferences described in \cite{Bewley86}.\footnote{Although we do not pursue it, a similar exercise could be performed in the multiple utility framework of \cite{Aumann62}.} This extension shows that our results are not limited to interval orders and the notions of upper and lower utilities, but apply to any situation in which rational behavior is driven by two distinct thresholds.

We apply our methods to precinct-level data from Lorain county in Ohio, focusing on the two races for the Ohio Supreme Court that occurred in the 2018 midterm elections, and demonstrate how several behavioral assumptions can be combined to reduce the size of the identification set. This empirical application fits our setting in several ways. First, individual choices are not observable. Second, the phenomenon of roll-off voting can be thought of as a channel through which incompleteness can directly manifest itself: more than 20\% of the turnout voters did not vote in these races even though they went to the polls.\footnote{A possible reason is that some voters cannot use party affiliation to rank candidates since it is not on the ballot even though candidates were selected by each party in their primary elections.} Third, voter registration data help us illustrate the role of limited consideration sets by assuming that partisan voters only consider candidates of their own party. Finally, Ohio election rules provide an example of an instrumental variable because the order in which the candidates are presented on the ballot changes from precinct to precinct. Order on the ballot is unlikely to be correlated with voters' preferences, while it can have an effect on vote shares, particularly when each candidate's party affiliation is not on the ballot as is the case for Judicial elections in Ohio. We show that the order has a noticeable, yet small, effect on the identification region in our races. 

\noindent \textbf{Literature Review}

Our paper brings together two strands of literature: the one focused on theories of individual decision making that yield indeterminacy in behavior, and the one focused on the econometric identification of economic models that have non-unique predictions. The former group dates back to \cite{Luce56}, who speaks about intransitive indifference, and has focused mostly on identifying individual preference characteristics from individuals choices (see \cite{Dziewulski-2021} for a recent example). To the best of our knowledge, we are the first to tackle econometric identification of incomplete preferences from choice data, including aggregate data, and we are the first to connect preference incompleteness in a deterministic setting with partial identification.

Our approach to the econometric identification of choice probabilities when preferences are incomplete follows the vast literature on partial identification. \cite{Manski2003} coins the term \textit{The Law of Decreasing Credibility} saying that ``\textit{The credibility of inference decreases with the strength of the assumptions maintained}''. Seeking more credible results, a researcher can either weaken assumptions on the behavior of decision makers in the model or weaken assumptions on the data generating process. Our main focus is the first of these two - the behavioral aspect, and the methods developed in this paper apply to any situation where decision makers cannot fully rank the alternatives they face.  

Ambiguity (or Knightian uncertainty) has generated much interest in the econometric literature on partial identification in the last two decades starting from \cite{Manski2000}. Ambiguity requires that (a) the analyst and the decision makers agree on the set of possible states of nature and (b) that the analyst knows the set of probabilities used by each decision maker (behavioral ambiguity) or knows the set that contains each decision maker's expectations (observational ambiguity). \cite{ManskiMolinari2008} show that questioner design can lead to imprecise knowledge of agents preferences.  \cite{GiustinelliManskiMolinari2021} elicit expectations from decision makers who may have imprecise subjective probabilities regarding late-onset dementia, and find that about half of the individual hold imprecise probabilities regarding their future health. \cite{Manski2018} looks at random utility models with linear utilities when some alternatives' characteristics are state dependent and the distribution function of the state is known to belong to a subset of the simplex. Section 2 of \cite{Manski2018} discusses observational ambiguity where agents have a unique subjective expectation for future states but the analyst only knows that this distribution belongs to a certain set of distributions. Section 3 of \cite{Manski2018}  discusses behavioral ambiguity when decision makers do not have a unique subjective distribution on the states of nature. In both cases Manski assumes that a decision maker does not chose an alternative which is dominated by another. He shows that in a binary choice model when choices are observed and the degree of ambiguity is known, this non-dominance condition yields inequalities on the parameters of the model. The results described in our paper generalize his findings to non-binary choice models with general behavioral incomplete preferences. We also combine the behavioral source of partial identification with observational source of partial identification.

Limited consideration sets are related to theories introduced in \cite{EliazSpiegler2011}, \cite{Masatlioglu2012}, and \cite{Lleras2017}. \cite{ManziniMariotti2014}) introduce assumptions on the way individuals restrict their attention to a subset of the alternatives such that the parameters of the model are point identified. When consideration sets are unobserved, the model parameters are partially identified. \cite{BarseghyanCoughlinMolinariTeitelbaum2021} consider decision makers with complete preferences but with unobserved consideration sets. Their model leads to partial identification of model parameters such as risk aversion. \cite{BarseghyanMolinariThirkettle2021} consider decision makers with unobserved heterogeneous consideration sets and use exclusion restriction on alternatives characteristics to point identify the parameters of the model. To a large extent the models of consideration sets constrain the behavior of the individuals rather than relax the assumptions of the model. In the application we consider, voting, the set of feasible alternatives is revealed on the ballot and is similar to all decision makers for statewide races. We analyze the impact of an assumption that certain voters consider only a subset of the candidates that are affiliated with a certain party.



Finally, this paper is also related to behavioral models where rationality is weakened. Some papers focus on finding tests for rationality in the general sense as in \cite{KitamuraStoye2018} and by \cite{Hoderlein2011}. These papers look for violations of utility maximizing behavior in data on individual choices. Violations, if found, are not attributed to failure of any specific assumption but rather to a general lack of rationality by some individuals. In our case, decision makers are perfectly rational even though their preferences are not complete.

\textbf{Structure}

The remainder of the paper is organized as follows. Section \ref{Incompleteness} introduces incomplete preferences and describes rational behavior. Section \ref{nonparIdentification} discusses random decision makers and non-parametric identification of preferences distribution. Section \ref{BinaryChoice} focuses on binary choice. Section \ref{extensions} illustrates several extension of our basic framework. In section \ref{Voting} we apply our findings to voting data, and show how this data can be used to illustrate our findings. Section \ref{conclustions} concludes.


\section{Incomplete Preferences and Rationality
\label{Incompleteness}}

Our objective is to analyze the standard discrete choice model without imposing the assumption that preferences are complete. In the standard model, completeness is reflected by the idea that choice between alternatives is driven by the utility assigned to each. This follows from the well known result that, with a finite number of alternatives, any complete and transitive preference relation has a utility function representing it. The utility function assigns a number to each alternative (its utility), and the ordering of these numbers is used to rank alternatives and thus to model choices. Without completeness this is no longer the case: one cannot necessarily assign a utility value to each alternative. 

We focus on a model in which each alternative is associated with two numbers, and these two numbers are used to determine the ranking between alternatives, if such a ranking exists. Let $X$ be the set of alternatives over which an individual's preference order $\succ$ is defined. For each $x \in X$ we assume there exist two real numbers $\uh(x)$ and $\ul(x)$, with $\uh(x) \geq \ul(x)$. We refer to these two numbers as an alternative's upper and lower utility respectively, and we use the word vagueness to describe the length of the utility interval $[\ul(x),\uh(x) ]$ associated to each $x$. Upper and lower utility can be used to describe the individual's preference ordering between any two alternatives, and this preference can then be used to describe her behavior when faced with a particular subset of feasible alternatives. First, for any $x,y \in X$ we say that $x$ is (strictly) preferred to $y$ if and only if $\bar{u}(y) < \underbar{u}(x)$. In words, one alternative is preferred if its lower utility exceeds the upper utility of the other. Two alternatives are not comparable when neither this inequality nor its opposite are satisfied (their utility intervals overlap). We say that two alternatives are indifferent if their upper and lower utilities are the same (their utility intervals are the same). 

Next, we describe how choices are made from a given set of feasible alternatives $ \mathcal{A} \subseteq X $ given the preferences described above. The main idea is that choice must allow for the possibility that some alternatives are not ranked. In particular, an alternative can be chosen from a set provided there is no other option in that set that is strictly preferred to it.\footnote{This weakens the standard revealed preferences conclusions as noted in \cite{Eliaz-Ok-2006}.} Suppose only two alternatives are feasible, then one of them can be chosen if it is either preferred to the other, or if it is not comparable to it. In other words, an alternative can be chosen as long as it is not dominated. In general, when there are more than two feasible possibilities in $\mathcal{A}$, an alternative can be chosen provided there is no other option in $\mathcal{A}$ that is strictly preferred to it. We formalize these ideas using the following definition.

\begin{definition}\label{non-dominated}
Given $ \mathcal{A} \subseteq X $, the subset of \textbf{non-dominated alternatives} is defined as 
$$M(\mathcal{A})=\left\{ a\in \mathcal{A} :\forall b\in \mathcal{A},\ \bar{u}(a)\geq \underbar{u}(b)\right\}.$$
\end{definition}
$M(\mathcal{A})$ is the set of all alternatives in $\mathcal{A}$ that are not (strictly) dominated by any element of $ \mathcal{A} $. When preferences are complete, the set $M(\mathcal{A})$ includes only alternatives that are indifferent to each other. Without completeness, the set $M(\mathcal{A})$ may include several incomparable alternatives. 

Let $y \in \mathcal{A}$ denote an observable choice from the feasible set $\mathcal{A}$; when this set is not a singleton the decision maker chooses an alternative from it randomly. Rationality means that dominated alternatives cannot be chosen even if behavior can be random because some alternatives are not comparable to each other.

\begin{definition}\label{rationality} 
A \textbf{rational} decision rule is a mapping $\psi:\mathcal{A} \rightarrow \Delta(\mathcal{A})$ such that $\psi(y)=0$ for any $ y \notin M(\mathcal{A})$.\footnote{$\Delta(A)$ denotes the simplex over the set $A$}
\end{definition}

In the case of incomplete preferences, rationality does not make unique predictions about behavior. When preferences are complete, typical assumptions imply that indifference is a zero probability event, and thus the analyst can treat $M(\mathcal{A})$ as a singleton.\footnote{This is usually done by assuming that utility functions are continuous.} Without completeness, however, assuming that indifference is a zero probability event does not necessarily imply that $M(\mathcal{A})$ is a singleton because incompleteness is not as knife-edge as indifference. One can rule out indifference between $x$ and $y$ by asking that the intervals $[\ul(x),\uh(x)]$ and $[\ul(y),\uh(y)$] are different, but this is not enough to rule out any overlap between them. Rationality also reflects the idea that no selection rule is a-priory imposed among incomparable alternative because it only says that an alternative that is strictly dominated is never chosen.

Although we cast the paper in the language of upper and lower utilities for ease of exposition, our identification results apply to any incomplete preference relation that yields a set of undominated alternatives constructed using only two numbers as stated in Definition \ref{non-dominated}. There are many examples of such preferences, and we end this section by describing one of them as an illustration of how upper and lower utilities can obtain; in Section \ref{sec:Knightian} we show that our approach is also consistent with the Knightian decision theory of \cite{Bewley86} where preferences are not complete because of ambiguity.\footnote{\cite{Bewley86} was later published as \cite{Bewley02}. Other examples could be \cite{Aumann62} and \cite{DubraMaccheroniOk04}, or the more recent \cite{Echenique-Pomatto-Vinson} and \cite{Miyashita-Nakamura}.}


\subsection{Interval Orders and Vagueness}
\label{Vaguness}

Interval orders are presented in \cite{Fishburn1970book} and are well suited to illustrate our setup. They can be obtained under simple assumptions on preferences. Let $(X,\succ)$ be a partially ordered set such that $X$\ is a finite set and $\succ$ is a strict preference relation over $X$. The first property is irreflexivity: $\lnot ( x\prec x)$ (where $\lnot $ denotes logical negation). The second property is a special form of transitivity: $x\prec y$ and $z\prec w$ $\Rightarrow $ $x\prec w$ or $z\prec y$.\footnote{One can easily verify that these two properties imply the usual transitivity.} Fishburn calls a preference that satisfies these two properties an \textbf{interval order}, and shows that $\succ$ can be represented using two functions as the following theorem illustrates.

\begin{theorem*}[Fishburn (1970)] \label{Fishburn1970}
If $\succ $ is an interval order and $X$ is finite, then there exists two functions $\underbar{u}:X\rightarrow \mathbb{R}$ and $\sigma :X\rightarrow \mathbb{R}$ with $\sigma (x)>0$ for all $x\in X$, such that%
\begin{equation*}
x\prec y\text{\qquad if and only if\qquad } \ul(x) +\sigma (x) <\ul(y).
\end{equation*}
\end{theorem*}

Thus, $y$ is preferred to $x$ if and only if the utility of $y$ exceeds the utility of $x$ by some strictly positive amount. One can think of each alternative as being associated with an interval on the real line. The lower bound of that interval represents its utility, while the width of that interval represents imprecision in that utility. In this spirit, Fishburn calls $\sigma $ the \emph{vagueness} function since it measures the amount of imprecision in the utility associated with each alternative. The name interval order follows from the observation that alternatives are compared using intervals: when $y$ is preferred to $x$ the interval $\left[\ul(y),\ul(y)+\sigma(y)\right]$ lies to the right of the interval $\left[\ul(x),\ul(x)+\sigma(x)\right]$. Interval orders are clearly not necessarily complete. When the two intervals overlap, $x$ and $y$ are not comparable. In this setting, although strict preference is transitive, non-comparability is not. In other words, $x$ may be not comparable to $y$, and $y$ maybe not comparable to $z$, but $z$ is strictly preferred to $x$. This idea is sometimes referred to as intransitive indifference.\footnote{Quoting from Fishburn: \textquotedblleft For example, if you prefer your coffee black it seems fair to assume that your preference will not decrease as $x$, the number of grains of sugar in your coffee, increases. You might well be indifferent between $x=0$ and $x=1$, between $x=1$ and $x=2$, ... , but of course will prefer $x=0$ to $x=1000$.\textquotedblright}

Interval orders easily map to our framework by letting $\uh(x)= \ul(x) +\sigma(x)$ so that each alternative is associated with a pair of numbers measuring the lower and upper bound of the alternative's `utility interval'. Using these two values, one can then talk about behavior when faced with a particular set of possibilities.\footnote{Using Fishburn's example, only cups of coffee which contains low amounts of sugar can be chosen. As soon as sugar content is high enough to make a cup strictly worse than a cup with no sugar at all, this cup is a dominated alternative. It, as well as any cup that contains more sugar, will not be chosen.} Inspired by Fishburn, we use the term vagueness to describe the difference between upper and lower utility of an alternative.


\section{Nonparametric Identification}
\label{nonparIdentification}

\textbf{Notation.} Throughout the paper we use capital Latin letters to denote sets and random sets. We use lower case Latin letters for random vectors and random functions. We use lower case Greek letters for parameter vectors and capital Greek letters for parameter sets. For a set $A \subset \Re^k$, $A^c$ denotes its complement. For a finite set $A$, $|A|$ denotes its cardinality. Scripted Latin letters are used for spaces or collections of similar objects.
\begin{Assumption}\label{probSpace}
Let $(I,\mathcal{F} ,\bfP)$ be a non-atomic probability space on which all random vectors, functions and sets are defined.\footnote{In case the analysis is conditioned on a $\sigma$-algebra $\mathcal{B}$ generated by a random variable $x$ (covariates), the probability space does not include any $\mathcal{B}$-atoms. The analysis and results in this paper can be written conditional on such a vector of covariates $x$. We avoid this extra notations for simplicity.} 
\end{Assumption}
 We use $i \in I$ to denote a random individual from the population $I$. For each $i\in I$ the set of feasible choices is $\mathcal{A}$, and for each $a\in \mathcal{A}$ decision maker $i$ has a utility interval $\left[ \ul_i(a) ,\uh_i(a) \right]$ as described in Section \ref{Incompleteness}.\footnote{We assume the set of alternatives is the same for all decision makers for simplicity. We explore the role of limited consideration sets in Section \ref{sec:attention_sets} } As usual, utility values are known to the individual but not to the analyst.

We treat $\ul:\mathcal{A} \rightarrow \mathbb{R}$ and $\uh:\mathcal{A} \rightarrow \mathbb{R}$ as two random functions such that for every $a\in \mathcal{A}$, $\bfP( \ul(a)\leq \uh(a)) =1$. As in standard discrete choice models, we rule out indifference by assuming that $\ul(a)$ and $\uh(a)$ are continuous random variables for every $a \in \mathcal{A}$. Our objective is to learn about choice probabilities using data on choices. For example, similarly to the case of complete preferences, we would like to estimate the probability that a random individual prefers alternative $a$ to all other alternatives, $\bfP \left( \ul(a) \geq \uh(b) \, \forall b \neq a \right)$. Denoting this probability as $\theta_a$, complete preferences and zero probability of ties imply that $\sum_{a \in \mathcal{A}}\theta_a=1$. Moreover, rationality (choosing the alternative that yield the highest utility) implies that the preference probabilities, $\theta_a$, and the observed choice probabilities, $P_a$, are synonymous. When preferences are not complete, however, the probability that alternatives $a$ and $b$ are incomparable but both are preferred to alternative $c$ can be positive. In what follows we formalize the difference between preference probabilities and choice probabilities in the context of incomplete preferences.

For an individual $i \in I$, let $M_i=M_i(\mathcal{A})$ be the set of alternatives that are not dominated,
\begin{align}
\begin{split}
\label{choiceSet}
M_i &=\{a \in \mathcal{A} : \nexists b \in \mathcal{A} \text{ such that } \ul_i(b)> \uh_i(a) \} \\
&= \{a \in \mathcal{A} : \max\limits_{b \in \mathcal{A}} \ul_i(b) \leq  \uh_i(a) \}.
\end{split}
\end{align}

We can describe $M$ as a mapping $M:I \rightarrow \mathcal{K}(\mathcal{A})$, where $\mathcal{K}(\mathcal{A}) $ is the set of all non-empty subsets of $\mathcal{A}$. Since $\mathcal{A}$ is finite, $M_i$ is non empty (a maximum exists) and $\mathcal{K}(\mathcal{A})$ contains compact sets. Since $\ul$ and $\uh$ are random variables, for all $A \in \mathcal{K}(\mathcal{A}) $, $\left\{ i: M_i \cap A \neq \emptyset \right\} \in \mathcal{F}$. Therefore, $M$ is a \emph{random set} (see Appendix \ref{RandomSets} for a summary of the definitions and tools of random set theory used in the body of the paper).

For every (non-empty) $A \in \mathcal{K}(\mathcal{A})$ define the probability that a decision maker's set of non-dominated alternatives equals $A$ as
$$\theta_A = \bfP(M = A).$$
This definition extends the standard concept of choice probabilities in models with complete preferences. Complete preferences and no ties mean that $\theta_A>0$ if and only if $|A|=1$ and $\theta_A=0$ otherwise. When preferences are not complete, however, there is a set $A$ of cardinality bigger than $1$ such that $\theta_A>0$. The collection $\mathbf{\theta} = \{ \theta_A \}_{A \in \mathcal{K}(\mathcal{A}) }$ describes all choice relevant parameters of the joint distribution of $\mathcal{U}=\{\ul(a),\uh(a) : a \in \mathcal{A} \}$. 

The vector of preference parameters $\theta$ satisfies the following properties:
\begin{enumerate}
    \item $\theta_A \geq0$ for $\forall A \in \mathcal{K}(A)$ and,
    \item $\sum_{A \in \mathcal{K}(\mathcal{A})} \theta_A =1$. 
\end{enumerate}
Therefore, after ordering these parameters in some way, the vector of choice parameters is an element of the simplex $\Theta =\Delta(\mathcal{K}(\mathcal{A}))$.\footnote{In section \ref{sec:abstention} we discuss abstaining where decision makers do not have to choose any alternative in $\mathcal{A}$. Here we assume that the choice probabilities sum to $1$ and therefore the choice parameters sum to $1$ as well.}
    
Before observing any data, we can only say that the vector of preference parameters lies in the simplex and the sum of any subset of choice parameters lies between $0$ and $1$  (See Figure \ref{fig:simplex} in the next Section for an example). 

Rationality (see Definition \ref{rationality}) implies that individual $i$'s choice, denoted $y_i$, is an element of the random set $M_i$. Without completeness, rationality allows for $y_i$ to be chosen from $M_i$ randomly. In the context of random set theory, this behavioral assumption translates to the following measurability assumption.
\begin{Assumption}[measurability]\label{measurability}
$y$ is a random variable which is a (random) selection of the random set $M$, $y \in Sel(M)$.\footnote{See Definition \ref{def:selectionSet} in Appendix \ref{RandomSets} for basic results on random sets.}
\end{Assumption}
A random sample of decision makers from the population amounts to drawing a random sample of their sets of non-dominated alternatives. When preferences are not complete there is an additional layer of randomness: a decision maker chooses an alternative from the set $M_i$ in a way that can be random.\footnote{Although we do not pursue that route formally, the econometric model could be augmented with a selection mechanism that completes the model. See Definition 2.4 in \cite{BMM2011}.}

A well known result from random set theory, Artstein's Lemma, connects the containment functional of the random set $M$ to the distribution function of a selection from that set; this connection is described by the Artstein's Inequalities. Theorem \ref{idRegionTheorem} shows there are restrictions that these inequalities impose on the preferences parameter $\theta$ even if preferences are not complete. 

\begin{theorem}\label{idRegionTheorem}
Under Assumptions \ref{probSpace} and \ref{measurability}, the identification region associated with the random set $M$ for the choice parameters is given by
\begin{equation}\label{idRegion}
\Theta^{I}=\{\theta \in \Theta : \sum_{A' \subset A} \theta_{A'} \leq \bfP(y \in A), \; \forall A\subset \mathcal{A} \}\text{.}
\end{equation}
\end{theorem}

\begin{proof}
Artstein's Lemma (see Theorem \ref{Artstein} in Appendix \ref{RandomSets}) implies that $y$ is a selection of $M$ if and only if for all $A\subset \mathcal{A}$, 
\begin{equation}
C_{M}(A)\leq \bfP (y\in A).  \label{containment_inequality}
\end{equation}
The choice probabilities on the right-hand side of inequality (\ref{containment_inequality}) are identified from the data for any $A \subset \mathcal{A}$. The containment functional (see Definition \ref{def:containmentFunctional} in Appendix \ref{RandomSets}) on the left-hand side of (\ref{containment_inequality}) depends on the unknown joint distribution of $\mathcal{U}$. Therefore, these inequalities impose restrictions on the choice parameters that depend on this unknown distribution. Computing the containment functional gives, 
\begin{align*}
C_{M}(A)& =\bfP (M\subset A) \\
& = \sum_{A' \subset A} \bfP(M=A') \\ 
& =\sum_{A' \subset A} \theta_{A'}.
\end{align*}
Given our knowledge of $\{\bfP(y\in A)\}_{A\subset \mathcal{A}}$, the identification set is as defined in equation (\ref{idRegion}).
\end{proof}

Artstein's inequalities imply that the set $\Theta^I$ is the sharp identification region for the parameter $\theta$. These inequalities are both sufficient and necessary for a parameter to be included in the identification set and hence $\Theta^I$ is the sharp identification set (see \cite{BMM2011} and \cite{BMM2012}). Therefore, even without completeness, the data imposes restriction on parameters of the joint distribution of preferences. The connection between the choice parameters and the order of the upper and lower utilities can be understood from the following relationship:
\begin{align*}
 \bfP(M \subset A) & =\bfP(M\cap A^c=\varnothing ) \\
   & =\bfP(\underset{k\in A^c}{\max } \, \uh(k)<\underset{a\in \mathcal{A}}{\max } \, \ul(a)) \\
 & =\bfP(\underset{k\in A^c}{\max } \, \uh(k)<\underset{a\in A}{\max} \, \ul(a))
\end{align*}
Therefore, restrictions imposed on the choice parameters can be translated into restrictions on the relative order of the upper and lower utilities.

If there exists $a \in A$ such that $0<\bfP (y=a) <1$, then $\Theta^{I}$ is a strict subset of $\Theta$ (this is trivially true when there are at least two alternatives). Theorem \ref{idRegionTheorem} shows that $\theta_a \leq \bfP (y=a)$ which is strictly less than $1$.\footnote{To simplify notation, for $A=\{a\}$, a singleton, we denote $\theta_a=\theta_{\{a\}}$.} In defining the set $\Theta$ this restriction on $\theta_a$ is not imposed and therefore $\Theta^I \subsetneq \Theta$.

Next, we show that all the inequalities in the definition of the identification region are potentially binding. Let, 
\begin{equation*}
\Theta^{1}=\{\theta \in \Theta :
\theta_a=C_{M}(\{a\})\leq \bfP (y=a) \, \forall a\in \mathcal{A} \}
\end{equation*}%
be the set of choice parameters that satisfy Artstein's inequalities for subsets $A$ with $|A|=1$. The following result gives conditions for this set to be larger than the identification set defined in Theorem \ref{idRegionTheorem}.

\begin{proposition}\label{thm:necessaryInequalities}
If $\exists A \subset \mathcal{A}$ such that $|A| \geq 2$ and $\theta_A>0$, then $\Theta^I \subsetneq \Theta^1$. 
\end{proposition}

\begin{proof}
$C_M(\{a\})=\theta_a \leq \bfP(y=a) \, \forall a \in A$, by Theorem \ref{idRegionTheorem}. Summing over $a \in A$, $\sum_{a \in A} \theta_{\{a\}} \leq \bfP(y \in A)$. This inequality is part of the definition of $\Theta^1$. Also, $C_M(A)=\sum_{A' \subset A}\theta_A =\sum_{a \in A}\theta_a+\sum_{A' \subset A : |A'| \geq 2}\theta_A \leq \bfP(y \in A)$, by Theorem \ref{idRegionTheorem}. The last inequality implies that $\sum_{a \in A}\theta_a \leq \bfP(y \in A) - \sum_{A' \subset A : |A'| \geq 2}\theta_A$. This inequality is not part of the definition of $\Theta^1$. By our assumption, $\sum_{A' \subset A : |A'| \geq 2}\theta_A >0$. Therefore, $\Theta^I$ includes at least one additional binding inequality which is not included in $\Theta^1$. 
\end{proof}

Proposition \ref{thm:necessaryInequalities} illustrates that inequalities involving sets with cardinality greater than 1 are potentially binding (in addition to inequalities related to subsets with cardinality 1). In terms of the capacity functional, the condition in Theorem \ref{thm:necessaryInequalities} can be replaced with $\exists A \subset \mathcal{A}$ such that $|A|>2$ and $C_M(A)>\sum_{a \in A} \theta_a$. 

\begin{example}
Suppose the alternatives set is $\mathcal{A}=\{a_0,a_1,a_2\}$. To simplify notation, let the choice parameters be $\theta = (\theta_0,\theta_1,\theta_2,\theta_{01},\theta_{02},\theta_{12},\theta_{012} )$ where $\theta_0=\theta_{\{a_0\}}$, $\theta_{01}=\theta_{\{a_0,a_1\}}$ and similarly for the other subsets. Let $p_j = \bfP(y=j)$ for $j=0,1,2$ be the choice probabilities. The following set of inequalities have to be satisfied by Theorem \ref{idRegionTheorem}:
\begin{align*}
    \theta_0 &\leq p_0 \\
    \theta_1 &\leq p_1 \\
    \theta_2 &\leq p_2 \\
    \theta_0+\theta_1+\theta_{01} &\leq p_0 + p_1 \\
    \theta_0+\theta_2+\theta_{02} &\leq p_0 + p_2 \\
    \theta_1+\theta_2+\theta_{12} &\leq p_1 + p_2 \\
    \theta_0+\theta_1+\theta_2+\theta_{01}+\theta_{02}+\theta_{12}+\theta_{012} &\leq p_0+p_1+p_2=1 \\
    \theta_0,\theta_1,\theta_2,\theta_{01},\theta_{02},\theta_{12},\theta_{012} & \geq 0.
\end{align*}
The last two inequalities define $\Theta$ (the simplex). The first three inequalities with the last two constitute $\Theta^1$, which uses only inequalities related to sets of cardinality $1$. Combining all the inequalities above yields the (sharp) identification region $\Theta^I$ defined in Theorem \ref{idRegionTheorem}. If either $\theta_{01}$, $\theta_{02}$, $\theta_{12}$ or $\theta_{012}$ are strictly positive, then as Theorem \ref{thm:necessaryInequalities} shows,  the identification set, $\Theta^I$, is a strict subset of $\Theta^1$. Moreover, it is clear from the above inequalities that the identification set is a convex subset of the simplex and therefore can be easily computed.
\end{example}

So far, we have obtained a general characterization of the identification region, and explored some of its properties. Next, we limit attention to a binary choice; this enables us to illustrate what we have learned so far, as well as present our next main results, using simple pictures and focusing on intuition.

\section{Binary Choice}
\label{BinaryChoice}

In this section we focus on binary choice models; we assume that the set of alternatives is $\mathcal{A}_i=\left\{ a_0,a_1\right\} $ for all $i$. As an illustration of Theorem \ref{idRegionTheorem}, we first derive the identification region for parameters of interest similar to those identified in models with complete preferences, and show that these parameters are only partially identified. Since there are only two alternatives, we can illustrate our results with diagrams. Then, we present the second main result of the paper: an instrumental variable can shrink the identification region and potentially rule out the hypothesis that all individuals have complete preferences. Finally, we study how the identification region is affected by abstention and limited consideration sets.

\subsection{No-assumptions Bounds}\label{noAssumptionsBounds}
We next derive the identification region implied by Theorem \ref{idRegionTheorem}, and show how it differs from the point identified case of complete preferences. We start with the case in which data identifies only the fraction of individuals who chose each alternative. We want to find out what these fractions can tell us about the distribution of utility values in the population. Formally, let $u_{i}(a_j) = [\ul_i(a_j) , \uh_i(a_j) ]$ for $j=0,1$; this is the utility interval agent $i$ assigns to alternative $a_j$. Let $\mathcal{U}=(\ul(a_0),\uh(a_0),\ul(a_1),\uh(a_1))$ be the corresponding random vector of utilities. We seek to identify, or partially identify, features (parameters) of the joint distribution of $\mathcal{U}$. For binary choice in particular, one is interested in the relative position of the utility intervals $[\ul(a_0),\uh(a_0)]$ and $[\ul(a_1),\uh(a_1)]$. 

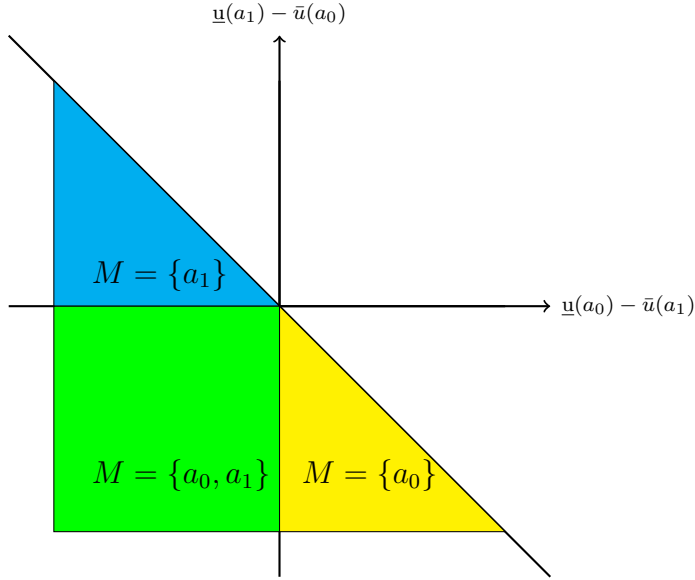
\begin{figure}[h!]
\centering
\begin{tikzpicture}[scale=3]
\draw [<->,thick] (0,1.2) node (yaxis) [above] {\scriptsize{$\ul(a_1)-\uh(a_0)$}} 
               |- (1.2,0) node (xaxis) [right] {\scriptsize{$\ul(a_0)-\uh(a_1)$}};
\draw [thick] (-1.2,0) -- (1,0);
\draw [thick] (0,-1.2) -- (0,1);
\draw [thick] (-1.2,1.2) -- (1.2,-1.2);
\filldraw[fill=yellow] (1,-1) -- (0,0) -- (0,-1) -- cycle;
\filldraw[fill=cyan] (-1,1) -- (0,0) -- (-1,0) -- cycle;
\filldraw[fill=green] (-1,-1) -- (-1,0) -- (0,0) -- (0,-1) -- cycle;
\filldraw[black] (0.05,-0.75) circle (0pt) node[anchor=west] {$M=\{a_0\}$};
\filldraw[black] (-0.88,0.14) circle (0pt) node[anchor=west] {$M=\{a_1\}$};
\node at (-0.15,-0.1) {(0,0)};
\filldraw[black] (-0.88,-0.75) circle (0pt) node[anchor=west] {$M=\{a_0,a_1\}$};
\end{tikzpicture}
\caption{Choice Rule with Incomplete Preferences} \label{fig:RandomSet}
\end{figure}

When preferences are complete, each interval is a singleton, and what matters for choice is the order of the utilities resulting from alternatives $a_0$ and $a_1$. Without completeness, an individual chooses from the random set of non-dominated alternatives, $M$, that is defined as 
\begin{equation*}
    M = \left\{ 
            \begin{array}{ll}
                        \{a_0\} & \text{if  }  \ul(a_0) > \uh(a_1) \\
                        \{a_1\} & \text{if  }  \ul(a_1) > \uh(a_0) \\
                        \{a_0,a_1\} & \text{otherwise.} 
                     \end{array}
                     \right.
\end{equation*}
Rationality means that $a_1$ is chosen with certainty if $\ul(a_1)>\uh(a_0)$, while $a_0$ is chosen with certainty if $\ul(a_0)>\uh(a_1)$.\footnote{Since we assume $\ul(a_i)$ and $\uh(a_i)$ are continuous random variables equality can be ignored.} Figure \ref{fig:RandomSet} illustrates the decision rule in terms of the random set $M$ and thus displays the decision maker's behavior. The axes are given by the difference between the lower utility of one alternative and the upper utility of the other, and the regions illustrating $M$ lie below the $-45^{\circ} $ line.\footnote{By definition, $\uh(a_j)-\ul(a_j)\geq 0$ for $j=0,1$; adding over $j$ and rearranging one gets $[\ul(a_1)-\uh(a_0)] \leq -[\ul(a_0)-\uh(a_1)]$. When upper and lower utilities coincide because preferences are complete, then $M$ coincides with the $-45^{\circ} $ line}

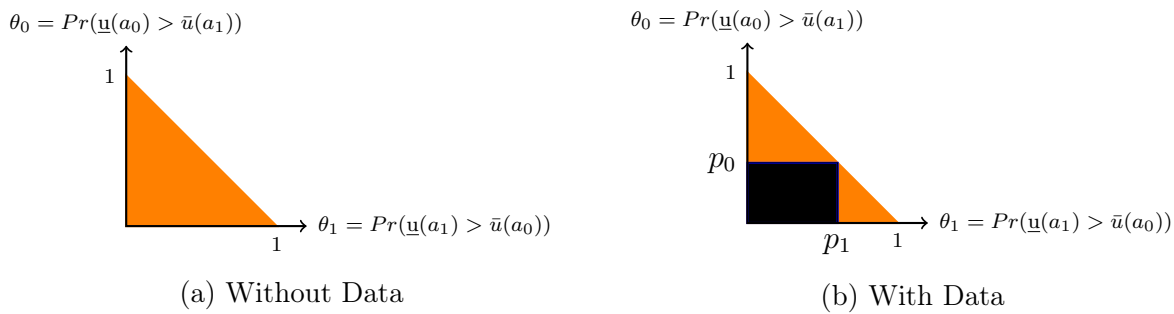
\begin{figure}
\begin{subfigure}{0.45\linewidth}
\begin{tikzpicture}[scale=2]

\filldraw[orange, thick] (0,0) -- (1,0) -- (0,1) -- cycle;
\draw (0,1) node[left] {{\scriptsize$1$}};
\draw (1,0) node[below] {{\scriptsize$1$}};

\draw [<->,thick] (0,1.2) node (yaxis) [above] {\scriptsize{$\theta_{0}=Pr(\ul(a_0)>\uh(a_1))$}} 
               |- (1.2,0) node (xaxis) [right] {\scriptsize{$\theta_{1}=Pr(\ul(a_1)>\uh(a_0))$}};

\end{tikzpicture}
\caption{Without Data}\label{fig:simplex}
\end{subfigure}
\hspace{6pt}
\begin{subfigure}{0.45\linewidth}

\begin{tikzpicture}[scale=2]

\filldraw[orange, thick] (0,0) -- (1,0) -- (0,1) -- cycle;
\draw (0,1) node[left] {{\scriptsize$1$}};
\draw (1,0) node[below] {{\scriptsize$1$}};

\filldraw[color=blue,fill=black, thick] rectangle (0.6,0.4);
\draw[color=black] (0,0.4) node[left] {$p_0$} -| (0.6,0) node[below] {$p_1$};

\draw [<->,thick] (0,1.2) node (yaxis) [above] {\scriptsize{$\theta_{0}=Pr(\ul(a_0)>\uh(a_1))$}} 
               |- (1.2,0) node (xaxis) [right] {\scriptsize{$\theta_{1}=Pr(\ul(a_1)>\uh(a_0))$}};

\end{tikzpicture}
\caption{With Data}\label{fig:partial}
\end{subfigure}
\caption{Binary Choice}
\end{figure}

Let
\begin{align*}
    \theta_0 &= Pr(\ul(a_0) > \uh(a_1)) \\
    \theta_1 &= Pr(\ul(a_1) > \uh(a_0))
\end{align*}
be the probabilities that a random decision maker prefers alternative $a_0$ over alternative $a_1$ and the probability that a random decision maker prefers alternative $a_1$ over alternative $a_0$, respectively. If preferences are complete and $\ul(a_j)=\uh(a_j)$ for $j=0,1$, then $(\theta_0+\theta_1)=1$. Due to incompleteness we can only say that $0 \leq \theta_0+\theta_1 \leq 1$. Without any data, there are no further restrictions. The identification region of $(\theta_0,\theta_1)$ before observing data is depicted in Figure \ref{fig:simplex} as the triangle between the axes and the negative $45^{\circ}$ line through the points $(0,1)$ and $(1,0)$. 

In our setting, aggregate choices are observed by the analyst and this data helps narrow down possible values of $(\theta_0,\theta_1)$. Let $y_{i}$ denote the choice made by individual $i$, and denote the choice probability of alternative $a_1$ by $p_1 = \bfP ( y_{i}=a_1)$ and the choice probability of $a_0$ by $p_0=\bfP ( y_{i}=a_0)$. These are the fractions of individuals that choose $a_1$ and $a_0$ respectively, and are identified from the data generating process. 

If preferences are complete, one has a point identified model where $(\theta_0,\theta_1)=(p_0,p_1)$. Without completeness, Theorem \ref{idRegionTheorem} implies that even if point identification is not possible partial identification is. An agent's choice, $y$, is a selection from the random set $M$ because of Assumption \ref{measurability}. Artstein inequalities imply that $y\in Sel(M) $ if and only if $\bfP(y \in K) \geq C_{M}(K) $ for every closed set $K$ where $C_{M}(K)$ is the containment functional. Substituting $K=\{a_0\}$ and $K=\{a_1\}$, this implies 
\begin{align*}
p_0=Pr(y \in \{a_0\}) \geq C_M(\{a_0\})=Pr(M \subset \{a_0\})=Pr(\ul(a_0)>\uh(a_1))=\theta_0 \\
p_1=Pr(y \in \{a_1\}) \geq C_M(\{a_1\})=Pr(M \subset \{a_1\})=Pr(\ul(a_1)>\uh(a_0))=\theta_1, 
\end{align*}
and thus the identification region is
\begin{equation}\label{equ:binaryIDregion}
    \Theta^I = \{ (\theta_0,\theta_1) \vert \  \theta_0 \leq p_0,  \ \theta_1 \leq p_1 \}. 
\end{equation}

Since all individuals who prefer $a_0$ choose it (and some individuals might choose it even though they cannot compare it with $a_1$), the probability that a randomly selected individual prefers $a_0$ cannot exceed the fraction $p_0$ of individuals who chooses it. If, for example, $p_{0}=0.37$ then the identification region cannot admit the case where $\theta_0 = 0.4$ and $\theta_1 =0.6$.

The size of the group of people who choose an outcome even though they cannot compare it to the other could go from zero up to all those who made that choice. For this reason, the discrete choice model that allows for incomplete preferences is partially identified. The dark area in Figure \ref{fig:partial} represents the identification region for $(\theta_0,\theta_1)$ given a pair of $p_0$ and $p_1$ values. The combinations of $( \theta_0, \theta_1 )$ which are included in the two light colored triangles in Figure \ref{fig:partial} are eliminated from the identification region after observing data. Observing choices informs the researchers about the possible values of the probability that $a_0$ is strictly preferred to $a_1$ and the probability that $a_1$ is strictly preferred to $a_0$.

The identification region in equation  (\ref{equ:binaryIDregion}) can be useful in understanding which features of the theoretical model can be deduced from data. For example, denote with $\theta_{01}$ the fraction of decision makers whose preferences are incomplete. Clearly, 
\begin{equation}
    \theta_{01} = 1-(\theta_0+\theta_1)
\end{equation}
where $\theta_0 +\theta_1$ is the proportion of decision makers who can compare the two alternatives. From Figure \ref{fig:partial} one notices that the possibility that $\theta_0 + \theta_1=1$ (and thus $\theta_{01}=0$) is included in the identification region: this is the point where the identification region touches the line connecting $\theta_0=1$ to $\theta_1=1$. Thus, one cannot rule out the possibility that all decision makers have complete preferences. Similarly, one cannot rule out the possibility that all decision makers cannot compare the two alternatives, so that $\theta_0 = \theta_1=0$ and $\theta_{01} = 1$ (this is the origin). One cannot state anything sharper than $0 \leq \theta_{01} \leq 1$ without additional information about preferences. 

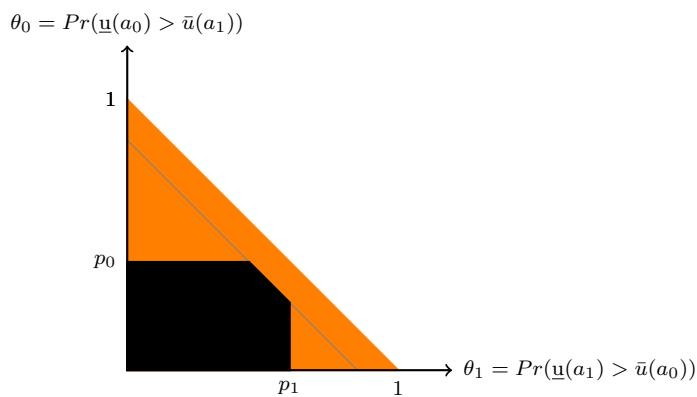
\begin{figure}[h!]
\centering
    \begin{tikzpicture}[scale=3.6]
        \filldraw[orange, thick] (0,0) -- (1,0) -- (0,1) -- cycle;
        \draw (0,1) node[left] {{\scriptsize$1$}};
        \draw (1,0) node[below] {{\scriptsize$1$}};
        \draw [gray,thin] (0.85 ,0) -- (0,0.85);
        \filldraw[color=black, fill=black, thick] (0,0) -- (0.6,0) -- (0.6,0.25) -- (0.45,0.4) -- (0,0.4) -- cycle;
        \draw [<->,thick] (0,1.2) node (yaxis) [above] {\scriptsize{$\theta_{0}=Pr(\ul(a_0)>\uh(a_1))$}} 
               |- (1.2,0) node (xaxis) [right] {\scriptsize{$\theta_{1}=Pr(\ul(a_1)>\uh(a_0))$}};
                \draw (0,1) node[left] {{\scriptsize$1$}};
        \draw (0.6,0) node[below] {{\scriptsize$p_1$}};
        \draw (0,0.4) node[left]  {{\scriptsize$p_0$}};

    \end{tikzpicture}
\caption{Partial identification with minimal amount of vagueness} \label{fig:minimal_vagueness}
\end{figure}

Suppose one knows that at least a proportion $\nu>0$ of decision makers cannot rank the two alternatives ($\theta_{01} \geq \nu$).\footnote{In Section \ref{Voting} data suggests that at least a certain proportion of voters may not have been able to rank the candidates.} Then $\theta_0 + \theta_1 = 1-\theta_{01} \leq 1- \nu$. The corresponding identification region is shown in Figure \ref{fig:minimal_vagueness}. The dark region includes all points consistent with two statements: at least $\nu$ proportion of decision makers cannot rank the two alternatives, and proportion $p_{0}$ of them chose alternative $a_0$.

Our results so far show that under reasonably few standard assumptions identification, albeit partial, is possible when preferences are not complete. Lack of completeness does not imply that ``anything goes''. Next, we introduce the concept of an instrumental variable, and show how such a variable could shrink the identification region in an interesting way. 

\subsection{Instrumental variables}\label{IV}

In this section we show how instrumental variables can refine the identification region established in Section \ref{noAssumptionsBounds}. In particular, we define an instrument as a random variable that influences choices while having (almost) no effect on preferences. When these instruments exist, they can be used to rule out the possibility that all decision makers have complete preferences. We present this result for the case of binary choice, but it extends to the general case.\footnote{One complication with more than two alternatives is that one may have partial identification even when preferences are complete.} Intuitively, if observed behavior changes with the realization of this random variable, then it must be the case that some individuals' behavior was not dictated by an actual ranking between the alternatives. In our voting application, a possible instrumental variable is represented by the order in which two candidates are presented on the ballot.

\subsubsection*{Perfect Instruments}

Think of a random variable that is independent of utilities but is correlated with choice. Intuitively, whenever a decision maker can rank the alternatives her choice cannot depend on the realized values of this random variable. However, when a decision maker cannot compare the alternatives her choice could depend on the random variable realizations (this dependence need not be deterministic). If observed \emph{aggregate} choices change with the realized value of the instrument it must be that \emph{some} decision makers were not able to rank alternatives. We formalize the idea of an instrument as follows.

\begin{definition}\label{def:IV}
  Let $Z$ be a random variable with a non-empty support $\mathcal{Z}$. For every $z \in \mathcal{Z}$ and  for $j=0,1$ let $p_{j|z}=\bfP(y=a_j|Z=z)$. Let, 
  \begin{equation*}
    \Delta_0=\sup_{z\in \mathcal{Z}}p_{0|z}-\inf_{z\in \mathcal{Z}}p_{0|z}.
\end{equation*}
We say that $Z$ is an \textit{instrumental variable} if (1)  $(\ul(a_j),\uh(a_j)_{j=0,1})$ are independent of $Z$, and  (2) $\Delta _{0}>0$. 
\end{definition}

Independence of $(\ul(a_j),\uh(a_j)_{j=0,1})$ and $Z$ corresponds to independence of the instrumental variable and the unobservable element of the model - the underlying utilities in our case. The validity of this condition, however, is driven by behavioral assumptions which are case specific and are not testable. $\Delta_{0}>0$ corresponds to the relevance of the instrumental variable - the outcome variable (i.e. the choice) is affected by $Z$. The validity of this condition in Definition \ref{def:IV} is testable.  Given both assumptions in Definition \ref{def:IV}, an instrumental variable can then be used to establish the result that some individuals' preferences must not be complete.

\begin{theorem}\label{binaryIVidentification}
 Let $Z$ be an instrumental variable. Then,%
\begin{equation}\label{binaryIV}
\theta_0 \leq \inf_{z\in \mathcal{Z}}p_{0|z}\ \text {\hspace{12pt} and\hspace{12pt}} \ \
\theta_1 \leq \inf_{z\in \mathcal{Z}}p_{1|z},
\end{equation}
and $\theta_{01} > 0$.
\end{theorem}

\begin{proof}
By the independence assumption, $\theta_{0|z}=\bfP(M={a_0}|Z=z)=\bfP(\ul(a_0)>\uh(a_1)|Z=z)=\bfP(\ul(a_0)>\uh(a_1))=\theta_0$ for all $z \in \mathcal{Z}$.  Artstein's inequalities applied for each $z \in Z$ as in the proof of Theorem \ref{idRegionTheorem} imply that $\theta_0 \leq p_{0|z}$ for every $z \in \mathcal{Z}$. Therefore, $\theta_0 \leq \inf_{z\in \mathcal{Z}}p_{0|z}$ and similarly $\theta_1 \leq \inf_{z\in \mathcal{Z}}p_{1|z}$, establishing equation (\ref{binaryIV}).
Using these results and the definition of $\theta_{01}$:
$$\theta_{01} = 1-(\theta_0+\theta_1) \geq 1- \left( \inf_{z\in \mathcal{Z}}p_{0|z} + \inf_{z\in \mathcal{Z}}p_{1|z} \right)= \sup_{z\in \mathcal{Z}}p_{0|z} - \inf_{z\in \mathcal{Z}}p_{0|z}=\Delta _{0}>0$$
as desired.
\end{proof}

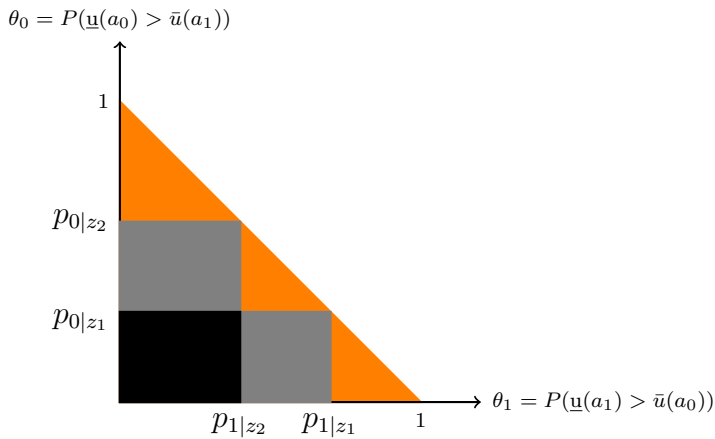
\begin{figure}[h!]
\centering
\begin{tikzpicture}[scale=4]

\filldraw[orange, thick] (0,0) -- (1,0) -- (0,1) -- cycle;
\draw (0,1) node[left] {{\scriptsize$1$}};
\draw (1,0) node[below] {{\scriptsize$1$}};

\draw [<->,thick] (0,1.2) node (yaxis) [above] {\scriptsize{$\theta_0=P(\ul(a_0)>\uh(a_1))$}} 
               |- (1.2,0) node (xaxis) [right] {\scriptsize {$\theta_1=P(\ul(a_1)>\uh(a_0))$}};
\filldraw[color=gray, fill=gray, thick] (0,0) -- (0.7,0) --  (0.7,0.3) -- (0,0.3) -- cycle;
\filldraw[color=gray, fill=gray, thick] (0,0) -- (0.4,0) --  (0.4,0.6) -- (0,0.6) -- cycle;
\filldraw[color=black, fill=black, thick] (0,0) -- (0.4,0) --  (0.4,0.3) -- (0,0.3) -- cycle;
\draw[color=black] (0,0.27) node[left] {$p_{0|z_1}$};
\draw[color=black] (0,0.6) node[left] {$p_{0|z_2}$};
\draw[color=black] (0.4,0) node[below] {$p_{1|z_2}$};
\draw[color=black] (0.7,0) node[below] {$p_{1|z_1}$};
\end{tikzpicture}
\caption{Identification Region with an Instrumental Variable} \label{fig:IDinstrumental}
\end{figure}

Theorem \ref{binaryIVidentification} says that the binding constraint on the probability that one outcome is preferred to the other is imposed by the lowest conditional probability that outcome is chosen, where conditioning is upon the values of the instrument. Figure \ref{fig:IDinstrumental} illustrates the identification power of having an instrumental variable. The identification region does not contain any point on the $-45^{\circ}$ line. Therefore, we can rule out the possibility that all individuals had complete preferences. The distance of the identification region from the $-45^{\circ}$ line depends on $\Delta_0$ which measures the extent to which choices are influenced by $Z$.


\subsubsection*{Imperfect Instruments}

One can, to a certain extent, relax the assumption that $Z$ is independent of the distribution of the utilities in $\mathcal{U}$ as explained in what follows. \cite{NevoRosen} introduced the notion of imperfect instrumental variables in a linear regression model with endogenous regressors. In their context, an imperfect instrumental variable is a variable $Z$ correlated with the error term of the regression but to a much lesser degree than it is correlated with the endogenous regressor. \cite{NevoRosen} show that, under some conditions, imperfect instruments can lead to partial identification of the regression parameters. We adapt the idea of imperfect instrumental variables to our model. Define the following quantities:
\begin{align*}
    \theta_{0|z} &=\bfP(\ul(a_0)>\uh(a_1)|Z=z) \\
    \theta_{1|z} &=\bfP(\ul(a_1)>\uh(a_0)|Z=z),
\end{align*}
and%
\begin{align*}
    \delta_0 =\theta_0 - \inf_{z \in \mathcal{Z}} \theta_{0|z} \\
    \delta_1 =\theta_1 - \inf_{z \in \mathcal{Z}} \theta_{1|z}.
\end{align*}
When Z is independent of preference parameters, as in Definition  \ref{def:IV}, $\delta_0 = \delta_1 = 0$ and one has a perfect instrument. If $Z$ is not a perfect instrument, $\delta_0$ and $\delta_1$ measure the sensitivity of the utilities to changes in the value of the instrument $Z$.

\begin{definition}\label{def:imperfectIV}

Let $Z$ be a random variable with a non-empty support $\mathcal{Z}$. For every $z \in \mathcal{Z}$ and  for $j=0,1$ let $p_{j|z}=\bfP(y=a_j|Z=z)$. We say $Z$ is an \textit{imperfect instrumental variable} if $\Delta _{0}>\delta_0+\delta_1$.

\end{definition}

The next result shows that if utilities depend on $Z$ to a lesser extent than choices depend on $Z$, the fraction of decision makers whose preferences are not complete is bounded away from $0$ by a strictly positive quantity.

\begin{theorem}\label{thm:imperfectIV}
If $Z$ is an imperfect instrumental variable, then $\theta_{01} >0$. 
\end{theorem}
\begin{proof}
As before, Artestein's inequalities imply that $\theta_{0|z} \leq p_{0|z}$ for all $z\in Z$; and therefore 
\begin{equation*}
\inf_{z \in \mathcal{Z}} \theta_{0|z} \leq \inf_{z\in \mathcal{Z}}p_{0|z}
\end{equation*}%
Using the definition of $\delta_0$
\begin{equation*}
\theta_0 -\delta_0 \leq \inf_{z\in \mathcal{Z}}p_{0|z}.
\end{equation*}%
 and similarly $\theta_1 -\delta_1 \leq \inf_{z\in \mathcal{Z}}p_{1|z}$. Summing these two inequalities gives 
\begin{equation*}
    (\theta_0+\theta_1) - (\delta_0+\delta_1) \leq \inf_{z\in \mathcal{Z}}p_{0|z} + \inf_{z\in \mathcal{Z}}p_{1|z}.
\end{equation*}
Since $\inf_{z\in \mathcal{Z}}p_{1|z} = 1- \sup_{z\in \mathcal{Z}}p_{0|z}$, we can write,
\begin{equation*}
    (\theta_0+\theta_1) - (\delta_0+\delta_1) \leq 1 - \Delta_0.
\end{equation*}
Finally,
\begin{equation*}
    \theta_{01} = 1 - (\theta_0+\theta_1) \geq \Delta_0 - (\delta_0 +\delta_1) > 0.
\end{equation*}
\end{proof}

\subsection{Abstention}\label{sec:abstention}

Next, we consider the possibility that choices are unobserved for a subset of decision makers. This situation can occur for several reasons. First, as is the case with voting (see Section \ref{Voting}), individuals can refrain from making a decision (abstention). Second, the data collected is incomplete due to non-response or non-observability. Let $p_0$, $p_1$, and $\gamma$ represent the proportion of decision makers who chose alternative $a_0$, alternative $a_1$ and abstained, respectively, such that $p_0+p_1+\gamma=1$.  Let $V \in \{0,1\}$ be a binary random variable indicating whether an individual's choice is observed, $V=1$, or unobserved, $V=0$. 

\subsubsection*{No assumptions bounds}
To identify the probability that a random individual strictly prefers $a_0$ over $a_1$ given that this decision maker's choice is observed one can use the methods described in Section \ref{noAssumptionsBounds}. For example, one could assume that voters that went to the polls and then abstained must have done so because their preferences were incomplete. However, we are interested in identification when such an assumption is not made. In particular, we want to identify $\theta_0$ and $\theta_1$ in the entire population of decision makers, not only among those who made a choice. Using the total law of probability, we can write
\begin{equation}\label{equ:abstain1}
    \begin{split}
    \theta_0 = \bfP(\ul_0>\uh_1) &= (1-\gamma)Pr(\ul_0>\uh_1|V=1)+\gamma Pr(\ul_0>\uh_1|V=0), \\
    \theta_1 = \bfP(\ul_1>\uh_0) &= (1-\gamma)Pr(\ul_1>\uh_0|V=1)+\gamma Pr(\ul_1>\uh_0|V=0).
    \end{split}
\end{equation}

\begin{figure}[h]
    \centering
    \begin{subfigure}{0.30\linewidth}
    \begin{tikzpicture}[scale=2]
        \filldraw[orange, thick] (0,0) -- (1,0) -- (0,1) -- cycle;
        \draw (0,1) node[left] {{\scriptsize$1$}};
        \draw (1,0) node[below] {{\scriptsize$1$}};
        \draw [<->,thick] (0,1.2) node (yaxis) [above] {\scriptsize{$\theta_{0}|Observed$}}
               |- (1.5,0) node (xaxis) [below] {\scriptsize{$\theta_{1}|Observed$}};
        \draw (0.625,0) node[below] {{\scriptsize$p_1$}};
        \draw (0,0.375) node[left]  {{\scriptsize$p_0$}};
        \filldraw[color=black,fill=black, thick] rectangle (0.625,0.375);
    \end{tikzpicture}
    \caption{Observed}
    \label{fig:abstaining-observed}
\end{subfigure}
\hspace{6pt}
\begin{subfigure}{0.30\linewidth}
    \begin{tikzpicture}[scale=2]
        \filldraw[black, thick] (0,0) -- (1,0) -- (0,1) -- cycle;
        \draw (0,1) node[left] {{\scriptsize$1$}};
        \draw (1,0) node[below] {{\scriptsize$1$}};
        \draw [<->,thick] (0,1.2) node (yaxis) [above] {\scriptsize{$\theta_{0}|Unobserved)$}}
               |- (1.55,0) node (xaxis) [below] {\scriptsize{$\theta_{1}|Unobserved$}};
    \end{tikzpicture}
    \caption{Unobserved}
    \label{fig:abstaining-Unobserved}
\end{subfigure}
\hspace{6pt}
\begin{subfigure}{0.30\linewidth}
    \begin{tikzpicture}[scale=2]
        \filldraw[orange, thick] (0,0) -- (1,0) -- (0,1) -- cycle;
        \draw (0,1) node[left] {{\scriptsize$1$}};
        \draw (1,0) node[below] {{\scriptsize$1$}};
        \filldraw[color=gray, fill=gray, thick] (0,0) -- (0.7,0) -- (0.7,0.3) -- (0.4,0.6) -- (0,0.6) -- cycle;
        \draw [<->,thick] (0,1.2) node (yaxis) [above] {\scriptsize{$\theta_{0}$}}
               |- (1.5,0) node (xaxis) [below] {\scriptsize{$\theta_{1}$}};
        \draw (0.4,0) node[below] {{\scriptsize$p_1$}};
        \draw (0.7,0) node[below] {{\scriptsize$p_1+\gamma$}};
        \draw (0,0.3) node[left]  {{\scriptsize$p_0$}};
        \draw (0,0.6) node[left]  {{\scriptsize$p_0+\gamma$}};
        \filldraw[color=black,fill=black, thick] rectangle (0.4,0.3);
    \end{tikzpicture}
    \caption{Joint}
    \label{fig:abstaining-joint}
    \end{subfigure}

    \caption{Partial identification with $\gamma$ abstaining}
    \label{fig:abstaining}
\end{figure}
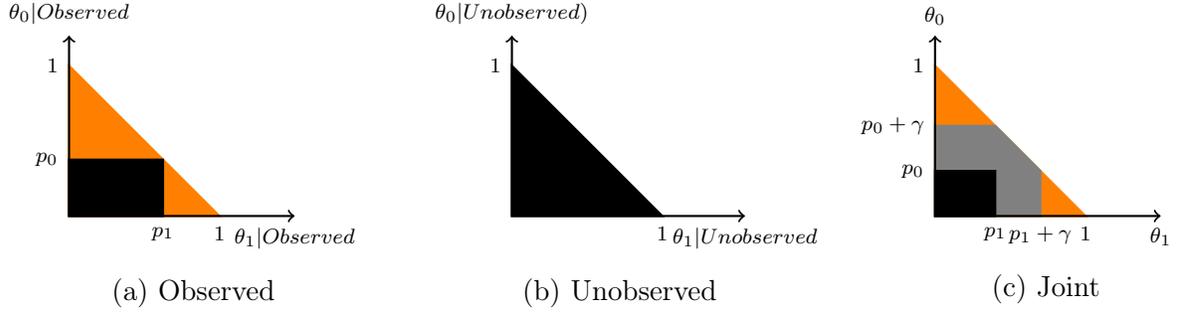

Rationality implies that
\begin{equation}\label{equ:boundsObserved}
    \begin{split}
        \bfP(\ul_0>\uh_1|V=1) \leq p_0,\\
        \bfP(\ul_1>\uh_0|V=1) \leq p_1.      
    \end{split}
\end{equation}
These inequalities are described by the rectangular identification region in Figure \ref{fig:abstaining-observed}. We denote this set as $\Theta^{O}=\{(\theta_0,\theta_1): 0 \leq \theta_0 \leq p_0 \, , \, 0 \leq \theta_1 \leq p_1  \}$.

The quantities $\bfP(\ul_0>\uh_1|V=0)$ and $\bfP(\ul_1>\uh_0|V=0)$ are unidentified and satisfy the following inequality 
$$0 \leq Pr(\ul_0>\uh_1|V=0)+Pr(\ul_1>\uh_0|V=0) \leq 1.$$

Finally, we can combine both identification regions $\Theta^O$ and $\Theta^U$ using
equation (\ref{equ:abstain1}),
\begin{equation}\label{equ:convexIDregion}
  \Theta^{I} =(1- \gamma) \Theta^O \oplus \gamma \Theta^U,  
\end{equation}
where $\oplus$ is the Minkowski sum. 

Since choices made by decision makers who abstain are unobserved, several assumptions can be made about $\Theta^U$. One can assume that all unobserved decision makers were not able to compare the two alternatives; in this case $\Theta^U=\{(0,0)\}$ and $\Theta^I = (1-\gamma) \Theta^O$ is the black rectangle in Figure \ref{fig:abstaining-joint}. Alternatively, one can be agnostic about the preferences of the unobserved decision makers and assume that $\Theta^U=\{ (\theta_0,\theta_1): 0 \leq \theta_0+\theta_1 \leq 1 \}$; these inequalities are represented by the triangular identification region in Figure \ref{fig:abstaining-Unobserved}. In this case the joint identification region is the gray polygon in Figure \ref{fig:abstaining-joint}. Another common  assumption is missing at random which implies that $\Theta^U=\Theta^O$ and yields $\Theta^I =\Theta^O$ which is the black rectangle in Figure \ref{fig:abstaining-observed}. Other assumptions on the behavior of the unobserved decision makers can be made depending on the application at hand.

\subsubsection*{Instrumental variables}

One can combine the ideas discussed above with the notion of instrumental variables. Let $Z$ be an instrumental variable with a finite support $\mathcal{Z}$ satisfying Definition \ref{def:IV}.
Equation (\ref{equ:abstain1}) holds for all values of the instrument $Z$. Furthermore, we assume that $\gamma$, the probability of abstaining remains constant across different values of the instrument $Z$. Equation (\ref{equ:boundsObserved}) holds for all values of the instruments which affect only the right hand-side. Therefore, 
\begin{equation}\label{equ:boundsObservedIV}
    \begin{split}
        \bfP(\ul_0>\uh_1|V=1) \leq \inf_{z \in \mathcal{Z}} p_{0|Z=z},\\ 
        \bfP(\ul_1>\uh_0|V=1) \leq \inf_{z \in \mathcal{Z}} p_{1|Z=z}.      
    \end{split}
\end{equation}

Combining equation (\ref{equ:abstain1}) and equation (\ref{equ:boundsObservedIV}), we have 
\begin{equation}
    \begin{split}
        \theta_0 \leq (1-\gamma) \inf_{z \in \mathcal{Z}} p_{0|Z=z} +\gamma \theta_{0|A=Unobs}, \\
        \theta_1 \leq (1-\gamma) \inf_{z \in \mathcal{Z}} p_{1|Z=z} +\gamma \theta_{1|A=Unobs}.
    \end{split}
\end{equation}
Combining this information with the fact that $0 \leq \theta_{0|V=0} +\theta_{1|V=0} \leq 1$, we can write the joint identification region similarly to equation (\ref{equ:abstain1}),
\begin{equation}\label{equ:convexIDregionIV}
    \Theta^{I|Z} =(1- \gamma) \Theta^{O|Z} \oplus \gamma \Theta^U,
\end{equation}
where $\Theta^{O|Z}=\{ (\theta_0,\theta_1): \theta_0 \leq \inf_{z \in \mathcal{Z}} p_{0|Z=z}. \, \theta_1 \leq \inf_{z \in \mathcal{Z}} p_{1|Z=z}  \}$. Theorem \ref{binaryIVidentification} implies that if $\sup_{z\in \mathcal{Z}}p_{0|z} - \inf_{z\in \mathcal{Z}}p_{0|z}>0$, then $\Theta^{O|Z} \subsetneq \Theta^O$ and therefore $\Theta^{I|Z} \subsetneq \Theta^I$.


Partial identification in this section is a combination of both behavioral assumptions (incomplete preferences) and observational challenges (abstention). The identification regions reported in equations (\ref{equ:convexIDregion}) and (\ref{equ:convexIDregionIV}) combine both sources of partial identification: $\Theta^U$ is the result of decisions makers whose choice is unobserved (e.g. voters abstaining); $\Theta^O$ and $\Theta^{O|Z}$ are a result of an incomplete model. The resulting identification regions $\Theta^I$ and $\Theta^{I|Z}$ are a convex combination of both sources of partial identification using the weight $\gamma$. \cite{Manski2018} discusses behavioral and observational sources of partial identification separately. Overall, combining both behavioral and observational reasons for partial identification have not received as much attention in the literature. Here, we showed how they can be combined and the application in Section \ref{Voting} illustrates this combination in practice.

\subsection{Limited Consideration Sets}\label{sec:attention_sets}
In Section \ref{IV} we show that instrumental variables allow the analyst to provide evidence that at least some decision makers have preferences that are not complete. The origin is included in all identification regions presented so far (see Figure \ref{fig:IDinstrumental}). Thus, one cannot rule out the possibility that all decision makers are unable to rank the two alternatives. In this section we explore a simple assumption that allows the analyst to exclude this possibility.

Consider the case where a known proportion of the decision makers considers, or pays attention to, only a subset of alternatives $\mathcal{A}' \subsetneq \mathcal{A}$.\footnote{See \cite{Masatlioglu2012} and \cite{Lleras2017} for formal models of decision making in the presence of limited consideration sets.} This could happen because some agents are unaware an alternative exist, or because some agents would never consider an alternative even when they are aware of it (for example, members of a political party would never vote for candidates who belong to a different party). When this happens, the set of non-dominated alternatives in Definition (\ref{non-dominated}) is applied to the decision maker with consideration set $\mathcal{A}'$ and gives the random set of non-dominated alternatives $M(\mathcal{A}')$.

\begin{figure}[h!]
    \centering
    \includegraphics[scale=0.65]{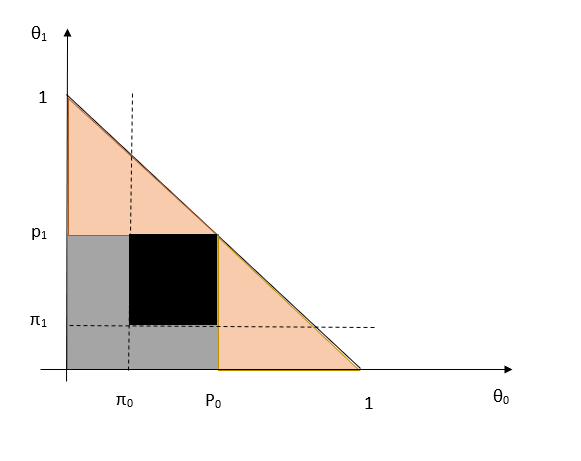}
    \caption{Partial identification with limited consideration sets}
    \label{fig:IDattentionSet}
\end{figure}

In a binary choice context $\mathcal{A}=\{a_0,a_1\}$, and there are only three potential consideration sets: $\mathcal{A}=\{a_0,a_1\}$, $\mathcal{A}_0=\{a_0\}$, and $\mathcal{A}_1=\{a_1\}$. Let $\pi_0$ be the fraction of decision makers who have consideration set $\mathcal{A}_0$ and $\pi_1$ the fraction that have consideration set $\mathcal{A}_1$, and assume that $\pi_0$ and $\pi_1$ are known to the analyst. Since decision makers with consideration set $\mathcal{A}_0$ do not consider, or are unaware of, alternative ${a_1}$ their choice of $a_0$ is deterministic from the analyst's standpoint. Similar reasoning applies to decision makers with consideration set $\mathcal{A}_1$. This model is coherent if $\pi_0 \leq p_0$ and $\pi_1 \leq p_1$. In this case
\begin{align*}
    \pi_0 &\leq \theta_0 \leq p_0, \\
    \pi_1 &\leq \theta_1 \leq p_1.
\end{align*}
If either $\pi_0$ or $\pi_1$ are strictly positive the identification set does not contain the origin. The case where $\pi_0,\pi_1>0$ is illustrated in Figure \ref{fig:IDattentionSet} where the identification set of $(\theta_0,\theta_1)$ is the darker rectangle. The combination of consideration sets and instrumental variables assumptions is demonstrated using our empirical application in Section \ref{Voting}.


\subsection{Parametric Linear Utility Model}\label{ParamtricLinearUtility}
Here we consider a simple parametric version of the binary choice model presented above as this is commonly used in applications. This version of the model helps us illustrate some features of the general model and is used in Section \ref{sec:ParametricPolicy}  to study the effect of policy interventions.

We focus on the case in which only the utility of one of the alternatives is interval-valued. In particular, utilities are as follows:
\begin{align*}\label{IntervalBinaryChoice}
    &\ul_i(a_0) = \uh_i(a_0) =0    \\
    &[\ul_{i}(a_1), \uh_{i}(a_1)] = [\beta_i - \sigma,\beta_i + \sigma].
\end{align*}
While the utility from alternative $a_0$ is assumed to be the singleton $0$ (location normalization), the utility from alternative $a_1$ is the random interval $[\beta_i-\sigma,\beta_i+\sigma]$. Note that $\sigma=0$ represents the case of complete preferences. Let $\beta_i =\betaBar - \varepsilon_i$ where $E(\varepsilon_i)=0$ and $Var(\varepsilon_i)=1$ (scale normalization). Assume that $\varepsilon \sim F_{\varepsilon}$ is continuously distributed and is independent across decision makers. For example, assuming that $\varepsilon_i$s are independent with a standard normal distribution yields the Probit model with incomplete preferences, while assuming a logistic distribution for $\varepsilon_i$ yields the Logit model with incomplete preferences. In most applications one assumes that $\beta_i=X_i \beta - \varepsilon_i$ where $X_i$ is the vector of observed characteristics of alternative $a_1$ as perceived by individual $i$. For simplicity we restrict our attention to case where no covariates are present.\footnote{\cite{Manski2018} discusses a similar model of linear utility functions in the context of Knightian uncertainty (see also our Section \ref{sec:Knightian} below). Moreover, estimation of this parametric model can be handled using modified minimum distance estimator developed in \cite{ManskiTamer2002}.}

Following equation (\ref{equ:binaryIDregion}), the identification region can be written as
\begin{align*}
   p_0 \geq \theta_0 =  \bfP(\ul_0>\uh_1) &=\bfP(\varepsilon_{i}> \beta +\sigma)  \\
    p_1 \geq \theta_1 = \bfP(\ul_1>\uh_0) &=\bfP(\varepsilon_{i}< \beta -\sigma)
\end{align*}
Combining these inequalities gives,
$$\bfP(\varepsilon_i< \beta -\sigma) \leq p_1 \leq \bfP(\varepsilon_i \leq \beta +\sigma).$$
If one assumes that $\varepsilon$ has a continuous everywhere monotone CDF, $F_{\varepsilon}$, the identification set is 
\begin{equation*}
\Theta^{I}=\left\{ \left( \beta,\sigma \right)\in \mathbb{R} \times \mathbb{R}_{+} :\beta -\sigma \leq F_{\varepsilon}^{-1}(p_1) \leq \beta +\sigma \right\}. 
\end{equation*}
Note that the above identification region includes the point where $\beta-\sigma=\beta+\sigma$ implying $\beta=F_{\varepsilon}^{-1}(p_1)$ and $\sigma=0$. In other words, the absence of vagueness cannot be excluded. 

In this simple setting, an instrumental variable is a random variable $Z$ such that the distribution of $\varepsilon|Z$ changes over the support of $Z$ while the determinants of utility, $\beta$ and $\sigma$, remain constant across the values of $Z$. If such a variable exists, the corresponding identification region is
\begin{equation*}
\Theta ^{I}=\left\{ \left( \beta ,\sigma \right) \in \mathbb{R} \times \mathbb{R}_{+}:\beta - \sigma \leq \inf_{z\in \mathcal{Z}}F_{\varepsilon}^{-1}\left(
p_{0}|z\right) \text{ and } \sup_{z\in \mathcal{Z}}F_{\varepsilon }^{-1}\left(
p_{0}|z\right) \leq \beta +\sigma  \right\} .
\end{equation*}
If an instrumental variable exists, one can reject the hypothesis that $\sigma=0$.


\section{Policy Intervention}\label{Policy}

In this section, we analyze the impact on the choice probabilities of a policy intervention aimed at changing the utility individuals give to one of the alternatives. Examples of such a policy could be campaign advertising, or other form of information provision, that targets a particular alternative. The objective of the policy is to increase the fraction of individuals who choose the targeted alternative. We show how this objective is harder to reach once incomplete preferences are allowed. We focus on binary choice and we first look at a model which is similar to the one of the previous Sections. We then specialize the model further and consider a simple linear utility parametric version of the general model. 

Formally, policy intervention is modeled by assuming one can add a positive quantity, $\Delta>0$, to the utility, both upper and lower, of alternative $a_1$. We denote by $p_1$ the choice probability of $a_1$ before the policy and by $p^{\Delta}_1$ the choice probability after adding $\Delta$ to the utilities from $a_1$. The effect of the policy is measured by $p^{\Delta}_1-p_1$. We first describe the policy effect when preferences are complete, and then move to the case in which completeness is not assumed.

\subsection{Non-parametric Policy Intervention}

Let $\mathcal{U}=(\ul_0,\uh_0,\ul_1,\uh_1)$ be the lower and upper utilities for alternatives $a_0$ and $a_1$, respectively. Except for assuming that $P(\ul_j \leq \uh_j)=1$ for $j=0,1$ we leave these utilities to be unspecified.

\subsubsection*{Complete preferences}
If individuals' preferences are complete, the utility of each alternative is a unique number, and therefore $\ul_0=\uh_0=u_0$ and $\ul_1=\uh_1=u_1$. In this case, $\theta_1$ and $\theta_0$ are point-identified by the corresponding choice probabilities. In particular, before the policy is introduced we have
\begin{equation*}
p_1=\theta_1=P(u_0<u_1)=P(u_0-u_1<0).    
\end{equation*}
After the policy intervention, the utility from choosing alternative $a_1$ is increased by $\Delta>0$, and becomes $u_1+\Delta$. In this case, the probability of choosing alternative $a_1$ must satisfy
\begin{equation*}
p^{\Delta}_1=\theta^{\Delta}_1=P(u_0<u_1+\Delta)=P(u_0-u_1<\Delta)
\end{equation*}
Clearly, $p^{\Delta}_1 \geq p_1$ and therefore one can predict that the policy impact, defined as $p^{\Delta}_1-p_1$, when preferences are complete is necessarily positive under some mild assumptions on the cumulative distribution function of $u_0-u_1$.
\begin{claim}
    Let $F()$ be the cumulative distribution function of $u_0-u_1$. If $F()$ is (strictly) monotonically increasing, then for $\Delta>0$, $p^{\Delta}_1-p_1$ is (strictly) positive. 
\end{claim}

\subsubsection*{Incomplete preferences}
Assume now that decision makers hold incomplete preferences. As before, let $p_1$ be the observed choice probability of $a_1$ before the policy intervention. We know from Section \ref{BinaryChoice} that this probability must satisfy the following inequality
\begin{equation*}
p_1 \geq \theta_1 = P(\ul_1 > \uh_0)= P(\ul_0 - \uh_1<0).
\end{equation*}
After adding $\Delta$ to both $\ul_1$ and $\uh_1$, the choice probability for $a_1$ will satisfy
\begin{equation*}
p^{\Delta}_1 \geq \theta^{\Delta}_1=P\left(\ul_0 - (\uh_1+\Delta)<0\right)=P(\ul_0 - \uh_1<\Delta).    
\end{equation*}
Again, we measure the policy impact by looking at $p^{\Delta}_1 - p_1$ after $p_1$ is observed.
\begin{claim}
    Let $F()$ be the CDF of $(\ul_0,\uh_1)$. Then, \\
    (1) the identification region of $p^{\Delta}_1 -p_1$ is 
    \begin{equation*}
        \Theta^{\Delta} = [1-p_1,F(\Delta)-p_1] 
    \end{equation*}
    (2) If $F()$ is strictly monotonically increasing and $\Delta < F^{-1}(p_1)$, then $0 \in \Theta^{\Delta}$.  
\end{claim}
Given the pre-policy choice probability and the fact that $p^{\Delta}_1 \geq F(\Delta)$, we can say that $p^{\Delta}_1 -p_1 \geq F(\Delta)-p_1$. The last difference can be negative if the policy intervention $\Delta$ is small enough to have $F(\Delta) < p_1$. Hence adding $\Delta>0$ to the utility from $a_1$ can lead to a decrease in the probability in which it is chosen - a result we could not get with complete preferences.

\subsection{Parametric Policy Intervention}\label{sec:ParametricPolicy}
Next, we add parametric assumptions as in Section \ref{ParamtricLinearUtility}. That is, utilities are:
\begin{align*}\label{IntervalBinaryChoice}
    &\ul_i(a_0) = \uh_i(a_0) =0    \\
    &[\ul_{i}(a_1), \uh_{i}(a_1)] = [\beta_i - \sigma,\beta_i + \sigma].
\end{align*}
and $\beta_i =\betaBar - \varepsilon_i$ where $\varepsilon$ has a standard normal distribution and therefore $\beta_i \sim N(\betaBar,1)$. Consider a policy intervention that has the ability to change the utility interval individual $i$ assigns to alternative $a_1$ by a fixed amount $\Delta>0$. After the policy intervention, the utility interval becomes
\begin{equation*}
u_i(a_1) = [\Delta + \beta_i -\sigma,\Delta + \beta_i+\sigma]. \label{equ:withPolicy}
\end{equation*}

\subsubsection*{Complete preferences}
When preferences are complete we have $\sigma=0$, and the choice probability for alternative $a_1$ before the policy is
\begin{equation*}
\Pr(y_i=a_1)=\Pr(\beta_i>0)=\Pr(\beta_i-\betaBar>-\betaBar)=1-\Phi(-\betaBar)=\Phi(\betaBar).    
\end{equation*}
Since $p_1 = \Pr(y_i=a_1)$ is identified from the data generating process, $\betaBar$ is identified as $\betaBar= \invPhi(p_1)$. After the policy is enacted, a quantity $\Delta>0$ is added to the utility of alternative $a_1$, and the choice probability for alternative $a_1$ is 
\begin{equation*}
    p^{\Delta}_1 = \Pr(\Delta+\beta_i>0) = \Pr(\beta_i- \betaBar > -\betaBar-\Delta)=1-\Phi(-\betaBar-\Delta)=\Phi(\betaBar+\Delta).
\end{equation*}
Since $\betaBar$ is identified, we can write, 
\begin{equation*}
    p^{\Delta}_1 = \Phi \left(\invPhi(p_1)+\Delta\right).
\end{equation*}
Therefore, the predicted effect of this policy is
\begin{equation*}
    p^{\Delta}_1 - p_1 = \Phi \left(\invPhi(p_1)+\Delta\right)-p_1>0.
\end{equation*}
because $\Delta>0$. Therefore, when preferences are complete the policy effect on the observed frequency with which alternative $a_1$ is chosen is always positive.

\subsubsection*{Incomplete preferences}
If preferences are not complete, $\sigma>0$ and the identification region is
$$\Theta^I = \{ (\Bar{\beta},\sigma) : \betaBar -\sigma \leq \invPhi(p_1) \leq \betaBar +\sigma \}.$$
The corresponding identification region for $\betaBar$ is
$$B^I = \left[\invPhi(p_1)-\sigma, \invPhi(p_1)+\sigma \right].$$
Therefore, observing $p_1$ does not point identify $\betaBar$ but rather gives a bound for it. A policy that adds $\Delta$ to the utility of alternative $a_1$ effectively changes $\betaBar$ to $\betaBar+\Delta$. Therefore, the predicted choice probability after the policy is
$$p_1^{\Delta} \in \{ p_1 : \Phi(\beta+\Delta-\sigma) \leq p_1 \leq \Phi(\beta+\Delta-\sigma), \ \beta \in B^I \}.$$
Here the effect of adding $\Delta$ to the utility of alternative $a_1$ is no longer guaranteed to be positive. The sign of the effect depends on the relationship between $\Delta$ and $\sigma$ as one can see in the diagrams in Appendix \ref{PolicyFigures}.

\section{Extensions}\label{extensions}

In this section we illustrate how some of our identification results can be adapted to additional assumptions about individuals' behavior or the data generating process. In particular, we discuss incompleteness due to ambiguity (Section \ref{sec:Knightian}) and minmax regret behavior (Section \ref{sec:minmax}). For ease of exposition, we focus on binary choice throughout.

\subsection{Discrete Choice With Knightian Uncertainty}\label{sec:Knightian}

Our identification framework can be adapted to different models of behavior when preferences are not complete as long as behavior is described by two thresholds. In the following, we illustrate how this could be done for Knightian uncertainty as described in \cite{Bewley02}. Bewley shows that a strict preference relation that is not necessarily complete, but satisfies all other axioms of a standard expected utility framework, can be represented by a family of expected utility functions generated by a unique utility index and a set of probability distributions. Lack of completeness is thus reflected in multiplicity of beliefs: the unique subjective probability distribution of the standard expected utility framework is replaced by a set of probability distributions. When the preference relation is complete this set becomes a singleton.

Let $S$ denote the state space and, with abuse of notation, its cardinality. $\Delta (S)$ is the set of all probability distributions over $S$. Given an alternative $x \in X\subset \boldsymbol{R^S}$, $u(x(s))$ denotes the utility that alternative yields in state $s$. If $\pi \in \Delta (S)$, the expected utility of individual $i$ according to $\pi$ is given by 
\begin{equation*}
E_{\pi }[u(x)] \equiv \sum_{s\in S}\pi (s) u(x,s).
\end{equation*}%
Let $\Pi \subset \Delta(S)$ be a closed and convex set of probability distributions on $S$.
According to Bewley's Knightian Decision Theory, decision maker $i$'s preferences $\succ $ are described by the following result: 
\begin{equation}\label{equ:Knightian}
x\succ y \qquad \text{if and only if}\qquad E_{\pi }[u(x)] >E_{\pi }[u(y)] \text{ for all }\pi \in \Pi.
\end{equation}%
 If the inequality in (\ref{equ:Knightian}) changes direction for different probability distributions in $\Pi$, the two alternatives are not comparable. 

This model does not compare alternatives using intervals. Even though the values of $E_{\pi }[u(\cdot)]$ form an interval, comparisons are not made looking at the extremes of that interval but one probability distribution at a time. Two alternatives could be ranked even if the corresponding utility intervals overlap. Despite this difference, the results of the previous section can be applied by taking advantage of some simple algebra. Rearranging equation (\ref{equ:Knightian}) one gets 
\begin{equation*}
x\succ y \qquad \text{if and only if}\qquad E_{\pi }[u(x)-u(y)] > 0 \text{ for all }\pi \in \Pi
\end{equation*}%
and therefore
\begin{equation*}
x\succ y \qquad \text{if and only if}\qquad \min_{\pi \in \Pi} E_{\pi }[u(x)-u(y)] > 0
\end{equation*}%

We can then adapt the definition of the set of non-dominated alternatives as follows 
\begin{equation*}
    M = \left\{ 
            \begin{array}{ll}
                        \{a_0\} & \text{if  }  \min_{\pi \in \Pi} E_{\pi }[u(a_0)-u(a_1)] > 0  \\
                        \{a_1\} & \text{if  }  \min_{\pi \in \Pi} E_{\pi }[u(a_1)-u(a_0)] > 0 \\
                        \{a_0,a_1\} & \text{otherwise.} 
                     \end{array}
                     \right.
\end{equation*}
As before, we let $\theta=(\theta_0,\theta_1)$ be the probabilities that alternative $a_0$ is preferred to $a_1$ and the probability that alternative $a_1$ ir preferred to $a_0$, respectively. By definition,
\begin{align*}
    \theta_0 &= \bfP(\min_{\pi \in \Pi} E_{\pi }[u(a_0)-u(a_1)] > 0) \\
    \theta_1 &= \bfP(\min_{\pi \in \Pi} E_{\pi }[u(a_1)-u(a_0)] > 0)
\end{align*}
From here on, the analysis can proceed along lines similar to the ones provided in Section \ref{BinaryChoice}, and we thus leave the details to the reader. A parametric version of this model is discussed in \cite{Manski2010} and \cite{Manski2018}. Manski presents the identification region for the parameters using two inequalities that corresponds to the Artstein's inequalities we use in this paper. For the binary choice, the inequalities in \cite{Manski2010} and \cite{Manski2018} are both necessary and sufficient to achieve sharpness. Generalizing this result to a multi-nomial choice model requires defining the random set $M$ of non-dominated alternatives similarly to equation (\ref{choiceSet}) and use Artstein's inequalities to form the sharp identification set (see \cite{BMM2011}). 

\subsection{Minmax regret}\label{sec:minmax}

So far, we have been silent about the way individuals make their choices when two or more alternatives are not comparable. This approach resulted in partial identification of the parameters of interest. Here we focus on the idea of minimizing maximal regret as a way to break the consumers' indecision. 

Assume that after a decision is made the individual learns which of the possible utility values represent her 'true' utility for each alternative. Regret occurs if the realized utility of the chosen option turns out to be lower than the realized utility of an alternative that was not chosen. For a binary choice situation, recall that 
\begin{equation*}
    M = \left\{ 
            \begin{array}{ll}
                        \{a_0\} & \text{if  }  \ul(a_0) > \uh(a_1) \\
                        \{a_1\} & \text{if  }  \ul(a_1) > \uh(a_0) \\
                        \{a_0,a_1\} & \text{otherwise.} 
                     \end{array}
                     \right.
\end{equation*}
and thus when the two utility intervals overlap both choices can be rational. If the individual chooses $a_0$ her maximal regret is $\uh(a_1)-\ul(a_0)$, while if she chooses $a_1$ her maximal regret is $\uh(a_0)-\ul(a_1)$. Therefore, if the decision maker minimizes her maximal regret when undecided, the corresponding rational choice region is as follows:
\begin{equation}\label{minmaxregret}
    M^{minmax} = \left\{ 
                     \begin{array}{ll}
                        \{a_0\} & \text{if \qquad}  \ul(a_0) - \uh(a_1) > 0 \text{\quad or \quad}  \ul(a_0) - \uh(a_1) > \ul(a_1) - \uh(a_0) \\
                        \{a_1\} & \text{if \qquad}  \ul(a_1) - \uh(a_0) > 0 \text{\quad or \quad}  \ul(a_0) - \uh(a_1) < \ul(a_1) - \uh(a_0) 
                     \end{array}
                     \right.
\end{equation}

\begin{figure}[h!]
    \centering
    \includegraphics[scale=0.6]{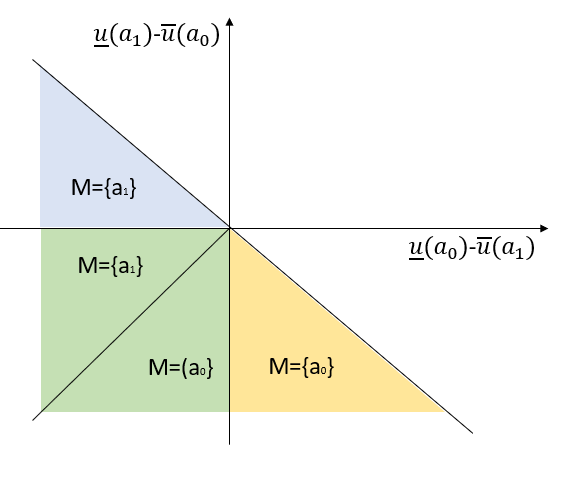}
    \caption{Choice rule with ex-post regret}
    \label{fig:choiceruleregret}
\end{figure}

The minmax regret rule selects one alternative from the random set $M$. Figure \ref{fig:choiceruleregret} describes the choice rule in equation (\ref{minmaxregret}). The region in Figure \ref{choiceSet} where choice was indeterminate is now split between alternatives so that points above the 45 degree line mean alternative $a_1$ is chosen and points below it mean alternative $a_0$ is chosen. Since choice is no longer indeterminate, the corresponding model is point identified.

\section{Illustration: Judicial Elections in Ohio}\label{Voting}

We implement the methods described in previous sections using data on voting.\footnote{Other papers estimated partially identified parameters in the context of voting. For example, \cite{KawaiWtanabe2013} estimate a model of strategic voting and quantify the impact it has on election outcomes. \cite{IaryczowerShiShum2018} estimate the effect of deliberation on collective choices in the context of criminal cases decided in the US courts of appeals. Both papers adopt set estimation as a result of multiple equilibria.} Specifically, we use precinct-level data obtained from the Ohio Board of Elections for the 2018 midterm elections in Lorain county, and focus on the two races for Justice of the Supreme Court of Ohio.\footnote{Since this paper deals with identification, we treat estimators as population level quantities and leave the statistical issues for future research.} We focus on one county to illustrate our results because it consists of a relatively homogeneous population. Election data fits our framework in many respects, the main one being that individuals' choices are not observable and one can only observe the share received by each candidate. The data from Ohio is also interesting because the candidates' order on the ballot varies from precinct to precinct. This change of order provides a potential instrumental variable. A full description of the data on Ohio 2018 midterm elections is in Appendix \ref{Data} including  explanation of how candidates' order on the ballot rotates. 

In the 2018 midterm elections voters in Ohio participated in eight statewide races. In six of these races candidates' party affiliation was listed on the ballot, while in the remaining two it was not. The races where no party affiliation was noted on the ballot were the two races for the position of Justice of the Supreme Court of Ohio. The candidates had been selected in party-run primaries, and thus were affiliated with a party, but their affiliation was not actually printed on the ballot itself. As Table \ref{tab:Lorain_overall} shows, in Lorain county the percentage of voters who refrained from expressing a preference is above 20\% when party affiliation is not listed. About 40,000 voters who went to the polls, voted in most state-wide races, did not vote in the two state-wide races where a candidate's party affiliation was not printed on the ballot.\footnote{Similar numbers hold statewide, with more than 800,000 voters not voting in the races for Justice of the Supreme Court after having gone to the polls.} This is an example of what is sometimes called roll-off voting: fewer votes are cast in down-ballot races. It represents a particular form of abstention because the voter has already incurred the costs associated with going to the polls.

\begin{table}[h]
\centering
\begin{tabular}{|l|c|c|c|}
\hline
                                 & Total Voters & {\% of Turnout} \\ \hline
Turnout                          & 116,231    &  100.00           \\ \hline
Governor and Lieutenant Governor & 114,551    &   98.55           \\ \hline
Attorney General                 & 110,236    &   94.84           \\ \hline
Auditor of State                 & 111,021    &   95.52           \\ \hline
Secretary of State               & 111,993    &   96.35           \\ \hline
Treasurer of State               & 110,919    &   95.43           \\ \hline
U.S. Senator                     & 113,855    &   97.96           \\ \hline
\rowcolor[HTML]{EFEFEF} 
Justice of the Supreme Court 1   &  87,525    &   75.30           \\ \hline
\rowcolor[HTML]{EFEFEF} 
Justice of the Supreme Court 2   &  85,472    &   73.54           \\ \hline
\end{tabular}
\caption{Lorain County Statewide Races}
\label{tab:Lorain_overall}
\end{table}

The recent literature on ballot roll-off in judicial elections (see \cite{Hall-Bonneau-2008} or \cite{Marble-2017}), is mostly focused on empirically measuring it and its possible determinants. Similarly to what we find, this literature shows that affiliation on the ballot decreases ballot roll-off. There is also a  theoretical literature explaining that abstention could stem from asymmetric information (\cite{Feddersen-Pesendorfer-1999}), or context-dependent voting (\cite{Callander-Wilson-2006}). In our framework, there could be an alternative reason for roll-off voting: incomplete preferences. When unable to rank alternatives, voters may decide not to make a choice and therefore do not vote in the corresponding race. Notice that a choice will be made for them anyhow, because the winner of the election will be selected as judge. In the context of voting, it is plausible to assume that those who came to the polls and did not vote behave that way because they could not compare the candidates. In the notation of equation \ref{equ:convexIDregionIV} in Section \ref{sec:abstention}, it means we can assume that $\Theta^U = \{(0,0)\}$. In other situations when a portion of the individuals do not report their choices (missing observations), a more plausible assumption is that $\Theta^U =\{(\theta_0,\theta_1):0 \leq \theta_0+\theta_1 \leq 1\}$.

The two races for a seat on the Ohio Supreme Court were Baldwin versus Donnelly and DeGenaro versus Stewart. As mentioned above, for these races the party affiliation of the candidates was not indicated on the ballot. We know, however, that Donnely, in the first race, and Stewart, in the second race, were affiliated with the Democratic Party. Table \ref{tab:Ohio_judges_races_Lorain} describes the results for these two races in Lorain county.

\begin{table}[h]
\centering
\begin{tabular}{|l|c|c|c|} 
\hline
                 & Baldwin votes & Donnelly votes & Total     \\ \hline
Justice of the Supreme Court 1          & 29,564     & 57,961      & 87,525 \\
                 & 33.8\%       & 66.2\%        &           \\ \hline
\hline 
                 & DeGenaro votes & Stewart votes & Total          \\ \hline
 Justice of the Supreme Court 2          & 37,282      & 48,190     & 85,472 \\
                 & 43.6\%        & 56.4\%       &           \\ \hline
\end{tabular}
\caption{\label{tab:Ohio_judges_races_Lorain} Ohio Supreme Court Races - Lorain County Results}
\end{table}

\begin{figure}[H]
\label{fig:empirical_noassumptions}
\begin{subfigure}[b]{0.35\linewidth}
\hspace*{-48pt}\begin{tikzpicture}[scale=3]

\filldraw[orange, thick] (0,0) -- (1,0) -- (0,1) -- cycle;
\draw (0,1) node[left] {{\scriptsize$1$}};
\draw (1,0) node[below] {{\scriptsize$1$}};

\draw [<->,thick] (0,1.2) node (yaxis) [above] {\scriptsize{$P(\ul(Baldwin)>\uh(Donnelly))$}} 
               |- (1.2,0) node (xaxis) [right] {\scriptsize {$P(\ul(Donnelly)>\uh(Baldwin))$}};
\filldraw[color=blue,fill=black, thick] rectangle (0.662,0.338);
\draw[color=black] (0,0.338) node[left] {\scriptsize$0.338$} -| (0.662,0) node[below] {\scriptsize$0.662$};
\end{tikzpicture}
\caption{Baldwin vs. Donnelly}
\label{fig:BaldwinDonnelly}
\end{subfigure}
\hspace*{36pt}
\begin{subfigure}[b]{0.35\linewidth}

\begin{tikzpicture}[scale=3]

\filldraw[orange, thick] (0,0) -- (1,0) -- (0,1) -- cycle;
\draw (0,1) node[left] {{\scriptsize$1$}};
\draw (1,0) node[below] {{\scriptsize$1$}};

\filldraw[color=blue,fill=black, thick] rectangle (0.564,0.436);
\draw[color=black] (0,0.436) node[left] {\scriptsize$0.436$} -| (0.564,0) node[below] {\scriptsize$0.564$};

\draw [<->,thick] (0,1.2) node (yaxis) [above] {\scriptsize{$P(\ul(DeGenaro)>\uh(Stewart))$}} 
               |- (1.2,0) node (xaxis) [right] {\scriptsize{$P(\ul(Stewart)>\uh(DeGenaro))$}};

\end{tikzpicture}
\caption{DeGenaro vs. Stewart}
\label{fig:DeGenaroStewart}
\end{subfigure}
\caption{Ohio's Supreme Court Races - 2018 Midterm Elections \\ No Assumptions Bounds }
\end{figure}
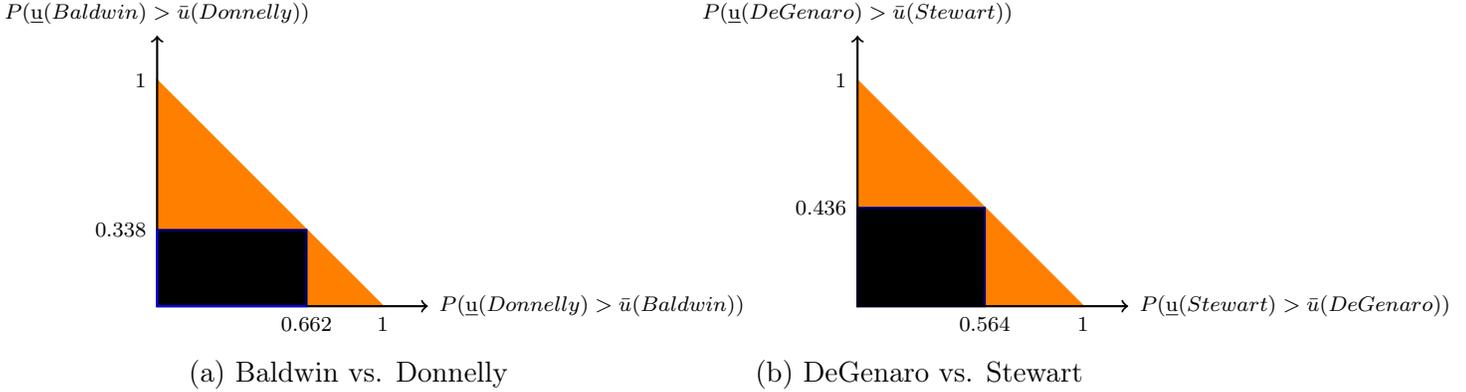

The no assumptions identification region for $(\theta_0,\theta_1)$ appears in Figures \ref{fig:BaldwinDonnelly} and \ref{fig:DeGenaroStewart}. These bounds are based on the choice probabilities of those who cast a vote in these races.

\subsection{Candidates Order}

The political science literature suggests that the order in which candidates are presented on the ballot may affect the chances of these candidates to be elected (see, for example, \cite{Krosnick-Miller-Tichy-04} and \cite{Meredith-Salant-2013}). As a result, ordering the candidates alphabetically may favor candidates with certain family names. For this reason, the Ohio Board of Elections rotates the order in which candidates appear on the ballot among the precincts (see Appendix \ref{Data}). As a result, a candidate may appear first on the ballot in one precinct and last in another nearby precinct. The order of a candidate on the ballot, by construction, presents an example of an instrumental variable. Focusing on Lorain County to achieve a relatively homogeneous population given the covariates, we use an instrumental variable indicating whether a candidate appeared first on the ballot at a certain precinct.

\begin{figure}[H]
\begin{subfigure}[b]{0.35\linewidth}
\hspace*{-48pt}\begin{tikzpicture}[scale=2.5]

\filldraw[orange, thick] (0,0) -- (1,0) -- (0,1) -- cycle;
\draw (0,1) node[left] {{\scriptsize$1$}};
\draw (1,0) node[below] {{\scriptsize$1$}};

\draw [<->,thick] (0,1.2) node (yaxis) [above] {\scriptsize{$P(\ul(Baldwin)>\uh(Donnelly))$}} 
               |- (1.2,0) node (xaxis) [right] {\scriptsize {$P(\ul(Donnelly)>\uh(Baldwin))$}};

\filldraw[color=gray,fill=gray] rectangle (0.675,0.325);
\filldraw[color=gray,fill=gray] rectangle (0.649,0.351);
\filldraw[fill=black] rectangle (0.649,0.325);

\draw[color=black] (0.790,0) node[below] {\scriptsize$0.675$};
\draw[color=black] (0.490,0) node[below] {\scriptsize$0.649$};

\draw[color=black] (0,0.395) node[left] {\scriptsize$0.351$};
\draw[color=black] (0,0.260) node[left] {\scriptsize$0.325$};

\end{tikzpicture}
\caption{Baldwin vs. Donnelly}
\label{fig:BaldwinDonnelly_IV}
\end{subfigure}
\hspace*{36pt}
\begin{subfigure}[b]{0.35\linewidth}

\begin{tikzpicture}[scale=2.5]

\filldraw[orange, thick] (0,0) -- (1,0) -- (0,1) -- cycle;
\draw (0,1) node[left] {{\scriptsize$1$}};
\draw (1,0) node[below] {{\scriptsize$1$}};
\draw [<->,thick] (0,1.2) node (yaxis) [above] {\scriptsize{$P(\ul(DeGenaro)>\uh(Stewart))$}} 
               |- (1.2,0) node (xaxis) [right] {\scriptsize{$P(\ul(Stewart)>\uh(DeGenaro))$}};

\filldraw[color=gray,fill=gray] rectangle (0.580,0.420);
\filldraw[color=gray,fill=gray] rectangle (0.547,0.453);
\filldraw[fill=black] rectangle (0.54,0.420);

\draw[color=black] (0.725,0) node[below] {\scriptsize$0.580$};
\draw[color=black] (0.400,0) node[below] {\scriptsize$0.547$};

\draw[color=black] (0,0.485) node[left] {\scriptsize$0.453$};
\draw[color=black] (0,0.350) node[left] {\scriptsize$0.420$};

\end{tikzpicture}
\caption{DeGenaro vs. Stewart}
\label{fig:DeGenaroStewartIV}
\end{subfigure}
\caption{Ohio's Supreme Court Races - 2018 Midterm Elections \\ Instrumental Variable }
\label{fig:Ohio IV}
\end{figure}
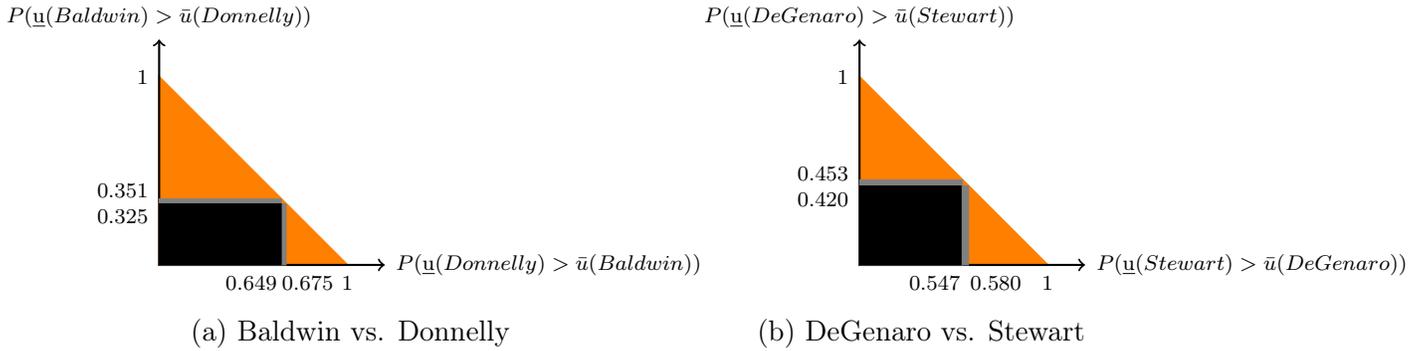

As one can see in equation (\ref{equ:CCP_order}), in both races for Ohio's supreme court judgeship, being first on the ballot gives a certain advantage over being second.\footnote{The advantage of appearing first on the ballot, when averaged over the whole state, is rather small. At the county level, the effect of the order is either large, rather small, or even negative. This uncounted difference between counties may be due to omitted covariates. We solve this issue by focusing on one county. Further empirical investigation is left for future work.} In the Baldwin versus Donnelly contest, appearing first on the ballot gives Donnelly a 2.6\% advantage on average. In the race of DeGenaro versus Stewart, the advantage of being first is estimated to be 3.3\% on average.  The predicted choice probabilities calculated at the mean values of the covariates are
\begin{align}\label{equ:CCP_order}
    P(y=Donnelly |county=Lorain,first=0) &=0.649 \notag \\
    P(y=Donnelly |county=Lorain,first=1) &=0.675 \\
    P(y=Stewart|county=Lorain,first=0) &=0.547 \notag \\
    P(y=Stewart|county=Lorain,first=1) &=0.580 \notag
\end{align}
and the corresponding identification regions are illustrated in Figure \ref{fig:Ohio IV}.

\subsection{Abstaining and Consideration Sets}
\begin{figure}[H]
\centering
\begin{tikzpicture}[scale=6]
    \draw [<->,thick] (0,1.25) node (yaxis) [left] {\scriptsize{$\theta_0$}} |- (1.25,0) node (xaxis) [below] {\scriptsize{$\theta_1$}};
    
    \draw[thick] (0.207,0)--(0.207,1.2);
    \draw[thick] (0,0.207)--(1.2,0.207);
    \draw[thick] (0,1)--(1,0);
    
    \filldraw[fill=gray,opacity=0.3] (0,0) -- (0,0.472) -- (0.528,0.472) -- (0.756,0.244) -- (0.756,0) -- cycle;
    \filldraw[fill=gray,opacity=0.3] (0,0) -- (0,0.505) -- (0.495,0.505) -- (0.737,0.263) -- (0.737,0) -- cycle;
    \filldraw[fill=orange, opacity=0.5] (0,0) -- (0.527,0) -- (0.527,0.244) -- (0,0.244) -- cycle;
    \filldraw[fill=orange,opacity=0.5] (0,0) -- (0.495,0) -- (0.495,0.263) -- (0,0.263) -- cycle;
    \filldraw[fill=black] (0.207,0.207) -- (0.495,0.207) -- (0.495,0.243) -- (0.207,0.244) -- cycle;
    
    \draw[color=black] (0,0.207)  -| (0.495,0) node[below] {\scriptsize{$0.495$}};
    \draw[color=black] (0,0.195) node[left] {\scriptsize{$0.207$}};
    
    \draw[color=black] (0,0.244)  -| (0.207,0) node[below] {\scriptsize{$0.207$}};
    
    \draw[color=black] (0,0.250) node[left] {\scriptsize{$0.244$}};
\end{tikzpicture}
\caption{Identification Set: Baldwin vs. Donnelly}
\label{fig:Empirical_BaldwinDonnelly}
\end{figure}
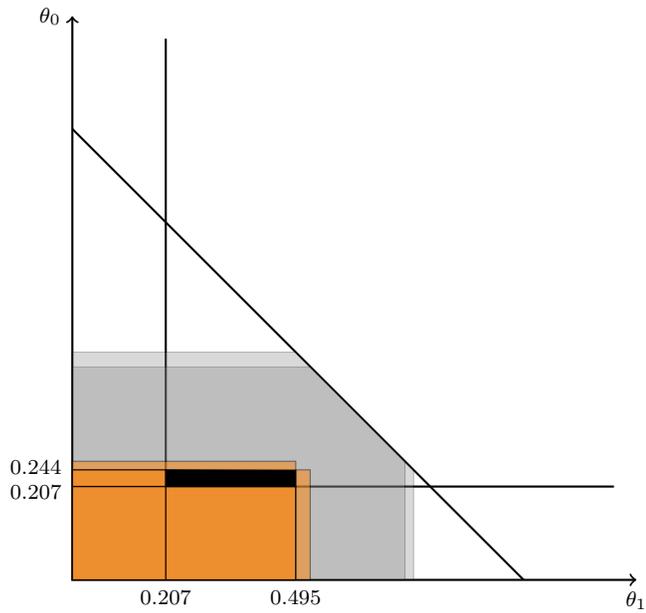

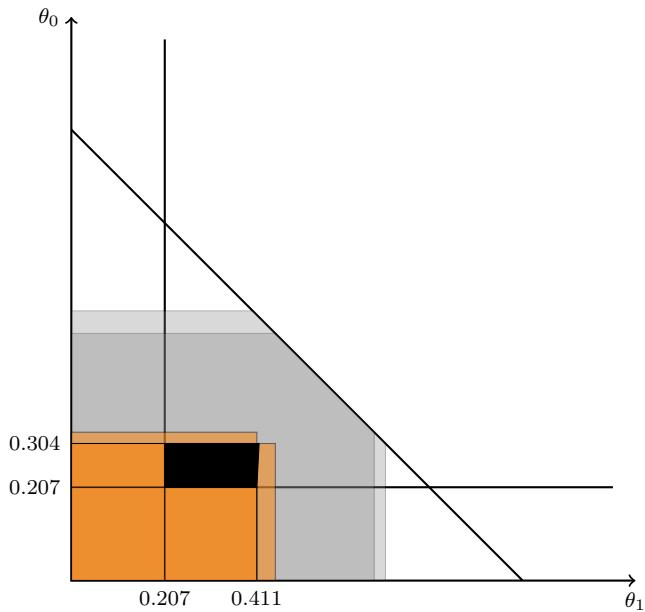
\begin{figure}[H]
\centering
\begin{tikzpicture}[scale=6]

    \draw [<->,thick] (0,1.25) node (yaxis) [left] {\scriptsize{$\theta_0$}} |- (1.25,0) node (xaxis) [below] {\scriptsize{$\theta_1$}};
    
    \draw[thick] (0.207,0)--(0.207,1.2);
    \draw[thick] (0,0.207)--(1.2,0.207);
    \draw[thick] (0,1)--(1,0);
    
    \filldraw[fill=gray,opacity=0.3] (0,0) -- (0,0.548) -- (0.452,0.548) -- (0.696,0.304) -- (0.696,0) -- cycle;
    \filldraw[fill=gray,opacity=0.3] (0,0) -- (0,0.598) -- (0.402,0.598) -- (0.671,0.329) -- (0.671,0) -- cycle;
    \filldraw[fill=orange, opacity=0.5] (0,0) -- (0.452,0) -- (0.452,0.304) -- (0,0.304) -- cycle;
    \filldraw[fill=orange,opacity=0.5] (0,0) -- (0.411,0) -- (0.411,0.329) -- (0,0.329) -- cycle;
    \filldraw[fill=black] (0.207,0.207) -- (0.411,0.207) -- (0.416,0.304) -- (0.207,0.304) -- cycle;
    \draw[color=black] (0,0.207) node[left] {\scriptsize{$0.207$}} -| (0.411,0) node[below] {\scriptsize{$0.411$}};
    \draw[color=black] (0,0.304) node[left] {\scriptsize{$0.304$}} -| (0.207,0) node[below] {\scriptsize{$0.207$}};
\end{tikzpicture}
\caption{Identification Set: DeGenaro vs. Stewart}
\label{fig:Empirical_DeGenaroStewart} 
\end{figure}

Next, we take into account that many voters have not expressed a preference in these races even though they have done so in other races. Following Equation (\ref{equ:convexIDregionIV}), we take a weighted average of the identification regions depicted in Figure \ref{fig:Ohio IV} and the identification region for those who did not vote. The weights, denoted as $\gamma$, are the conditional abstaining probabilities computed for each race. Specifically, using Table \ref{tab:Lorain_overall}, in the first race $\gamma=0.2470$ and $\gamma=0.2646$ in the second race. In this voting application, it is plausible to assume that all voters who abstained from choosing a supreme court judge could not compare the two candidates. This assumption corresponds to assuming that $\Theta^U=\{(0,0)\}$ as discussed in Section \ref{sec:abstention}. With this assumption the identification region corresponds to the orange rectangles in Figures \ref{fig:Empirical_BaldwinDonnelly} and \ref{fig:Empirical_DeGenaroStewart}. In other applications, however, it may be more plausible to assume that $\Theta^U=\{(\theta_0,\theta_1): 0 \leq \theta_0+\theta_1 \leq 1\}$.  The identification regions in this case are the gray areas in Figures \ref{fig:Empirical_BaldwinDonnelly} and \ref{fig:Empirical_DeGenaroStewart}.

Finally, we introduce consideration sets by assuming that registered democrats and republicans only considered voting for a candidate of their own party. Figures \ref{fig:Empirical_BaldwinDonnelly} and \ref{fig:Empirical_DeGenaroStewart} combine instrumental variables, consideration  sets, and the presence of abstention. The dark identification region uses as lower bound the fraction of voters who are registered for the candidate's party and as upper bound the conditional probabilities in equation (\ref{equ:CCP_order}) times the corresponding $1 - \gamma$.

The identification sets described in Figures \ref{fig:Empirical_BaldwinDonnelly} and \ref{fig:Empirical_DeGenaroStewart} demonstrate how all the tools developed in this paper can be combined together. In practice, however, a researcher could start from the no-assumptions bounds described in Section \ref{noAssumptionsBounds} and then add the assumptions she deems plausible for the application at hand - either all of them or only a subset. 

\section{Conclusions}\label{conclustions}
In this paper we provided the sharp identification region for a discrete choice model in which individuals' preference may not be complete and only aggregate choice data is available. The identification region is a strict subset of the parameter space, and thus disproves the idea that everything goes when preferences are not complete. The identification region provides intuitive bounds on parameters of interest of the probability distribution of preferences across the population. Since our assumption do not rule out complete preferences, the identification region admits the two extreme possibilities of maximal incompleteness in which nobody can rank alternatives and no choice-relevant incompleteness in which everyone can rank alternatives. We illustrate how the existence of instrumental variables can rule out this last possibility.

Although we use interval orders as a way to describe preferences and behavior when incompleteness is allowed, our results extend beyond that model. In particular, any theory where behavior depends on two thresholds (instead of just one) can be accommodated into our framework. We have illustrated this using the Knightian decision theory of \cite{Bewley02}, but we believe our results would also extend to the multi-utility framework of \cite{Aumann62} and \cite{DubraMaccheroniOk04}, or to the more recent twofold conservatism model of \cite{Echenique-Pomatto-Vinson}, as well as individual decision making models with ``thick'' indifference curves. All that one needs is to have a set of non-dominated alternatives that depends on two numbers per each alternative.

\newpage 
\appendix

\appendixpage

\section{Random Sets}
\label{RandomSets}
Identification analysis in economics has greatly benefited from tools developed in Random Set Theory (RST).\footnote{\cite{Molchanov2005} presents a general exposition of RST. We focus on real-valued random variables and sets. The reader is referred to Appendix A of \cite{BMM2012} and to \cite{Molchanov2005} for a more in depth discussion of RST.} We use results developed in \cite{BMM2011} and \cite{BMM2012} to identify the quantities and parameters of interest. Specifically, we focus on the containment functional approach to partial identification.\footnote{Another approach is called the Aumann Expectation approach to partial identification. \cite{BMM2012} discuss the merits of both approaches and make recommendations as to where each approach may have an advantage.} We start with defining random sets and then the concepts related to the two identification strategies that are used in later sections of this paper. 

The probability space, $(I,\mathcal{F} ,\bfP)$, on which all random variables and sets are defined is non-atomic. We use $i \in I$ to denote a random individual from the population $I$. From here on equalities and the statement "for every $i$" mean for every $i\in I$, $\bfP$-a.s. A random set is a measurable map defined as follows.

\begin{definition}\label{def:RandomSet}
A random set $X$ is a mapping $X:I \rightarrow \mathcal{K}\left(\mathbb{R}^{n}\right) $ where $\mathcal{K}\left(\mathbb{R}^{n}\right) $ is the set of all closed subsets of $\mathbb{R}^{n}$ and such that for all $K\subset \mathcal{K}\left(\mathbb{R}^{n}\right) $ compact, $\left\{ i:X\left( i\right) \cap K\neq \emptyset \right\} \in \mathcal{F}$.
\end{definition}

A random set can be thought of as a collection of point-valued random variables. The collection of all (point-valued) random variables, $x$, defined on our probability space such that $x\left( i\right) \in X\left(i\right) $ for all $i$ is defined as follows. 

\begin{definition}\label{def:selectionSet}
For a random set $X$, a selection of $X$ is a random variable $x$ such that $x(i) \in X(i)$ for all $i$. We let $Sel\left(X\right)$, the \emph{selection set} of $X$, be the collection of all selections of $X$.
\end{definition}

$Sel(X)$  is non-empty, see Theorem 2.13 in \cite{Molchanov2005}.

\subsection{Containment Functional}
The containment functional of a random set corresponds to the distribution function of a regular random variable.

\begin{definition}\label{def:containmentFunctional}
For a random set $X$ and $\forall K\in \mathcal{K}\left( \mathbb{R}^{n}\right) $, the containment functional is defined as%
\begin{equation*}
C_{X}\left( K\right) =\bfP \left( X\subset K\right) .
\end{equation*}
\end{definition}

The following result, sometimes referred to as Artstein's Lemma,\footnote{See \cite{Artstein1983}, \cite{Molchanov2005} Theorem 2.20 and Theorem 2.1 in \cite{BMM2012}.} establishes a relationship between the selection set and the containment functional.

\begin{theorem}\label{Artstein}
(Artstein's Inequalities) Let $X$ be a random set and let $Sel\left( X\right) $
be its selection set. Then $x\in Sel\left( X\right) $ if and only if 
\begin{equation*}
C_{X}\left( K\right) \leq \bfP \left( x \in K\right)
\end{equation*}%
for all $K\in \mathcal{K}\left( \mathbb{R}^{n}\right)$.
\end{theorem}

When a single selection $x$ from the random set $X$ is observed, $\bfP(x \in K)$ is identified from the data. It can then be used to draw restrictions on the possible values of the containment functional. This approach is especially useful when one looks at projections of the random set and its selections on a lower dimensional space as we show in section \ref{nonparIdentification}.


\section{Data}\label{Data}
The data used in Section \ref{Voting} on 2018 midterm elections in Ohio  was collected from the Ohio Board of Elections.\footnote{Most of the information is accessible from the Ohio Board of Elections web page at https://www.boe.ohio.gov. Additional information was obtained from county specific webpages, for example https://www.voteloraincountyohio.gov/2018-general-election for Lorain county.} The data contains information on all the races including the positions, the names of the candidates, number of registered voters, votes cast for each candidates and the order in which the candidates appeared on the ballot. In cases where the party affiliation of the candidates was revealed on the ballot, we collected this information as well. Over all there were 8904 precincts in Ohio in the 2018 midterm elections. Some ballots included up to twenty different races depending on the district in which the precinct is located. 

Midterm elections happen in the US every four years in between the General (presidential) elections. In general, midterm elections include three type of races; (1) National races (e.g. US senators and US congressmen), (2) State Races (e.g. governor and state supreme court judges) and (3) district/local races (e.g. state congress and board of education).

National races for positions like the governor of the state or the US senator are high profile races. The candidates in these races are affiliated with one of the two major parties - Democratic or Republican - and are well funded. The candidates use this money to widely advertise themselves and enjoy the support of their parties. As a result, voters are likely to be familiar with the names of these candidates when they come to vote. Possible exceptions are candidates who run for either the Green party or the the Libertarian party who are less widely known.

On the ballot there are also races who receive less attention in the media. These positions, like auditor of the state, do not receive the same level of attention and campaign funding. As a result, the candidates running for these positions are less known. Candidates for these positions are affiliated with a party and their affiliation is denoted on the ballot. Ohio does not allow straight ticket party voting.

Finally, there are races where candidates do not have party affiliation. For example, state supreme court judges and state board of education. In these races, candidates tend to be both less familiar to the voter and cannot enjoy party affiliation or their party affiliation is not indicated on the ballots.

The order in which candidates appear on the ballot is as follows. Within each county the precincts are ordered by the precinct's code. In the first precinct candidates appear on the ballot by their alphabetical order. Then in the next precinct and on the candidate that appeared last in the previous precinct on moves to first place on the ballot and the other candidates move one spot down each.

\section{Figures About Policy Effects}\label{PolicyFigures}

In Section \ref{Policy} we described how after a policy $\Delta>0$ is enacted, the predicted choice probability for alternative $a_1$ is given by 
$$p_1^{\Delta} \in \{ p : \Phi(\betaBar+\Delta-\sigma) \leq p \leq \Phi(\betaBar+\Delta-\sigma), \ \beta \in B^I \}.$$
Therefore, the effect of adding $\Delta$ to the utility of alternative $a_1$ can be positive or negative as one can see in the following diagrams.

\begin{figure}[ht]
\includegraphics[scale=0.53]{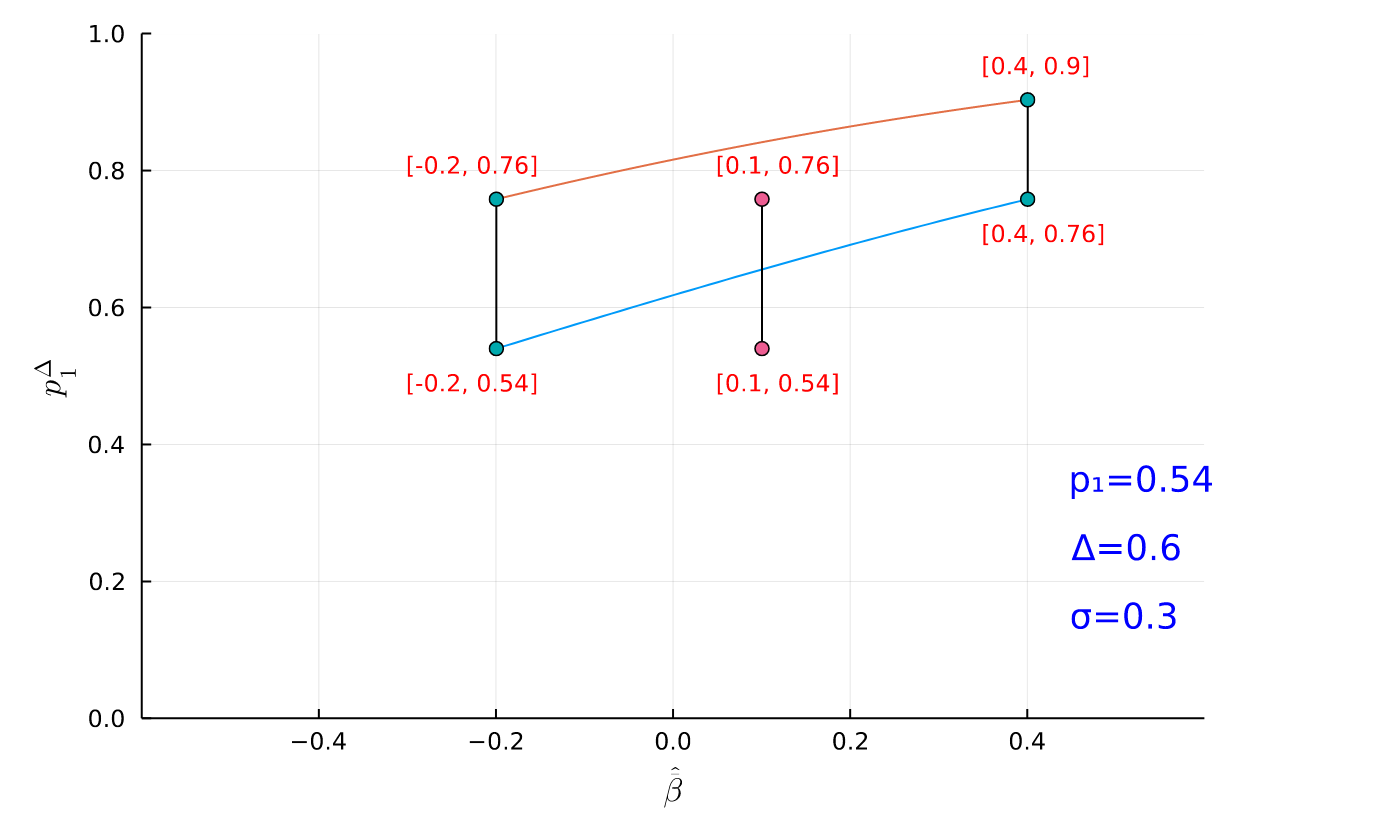}
\centering
\caption{Identification region for $p_1^{\Delta}$ when $2\sigma=\Delta$}
\end{figure}

\begin{figure}[ht]
\includegraphics[scale=0.53]{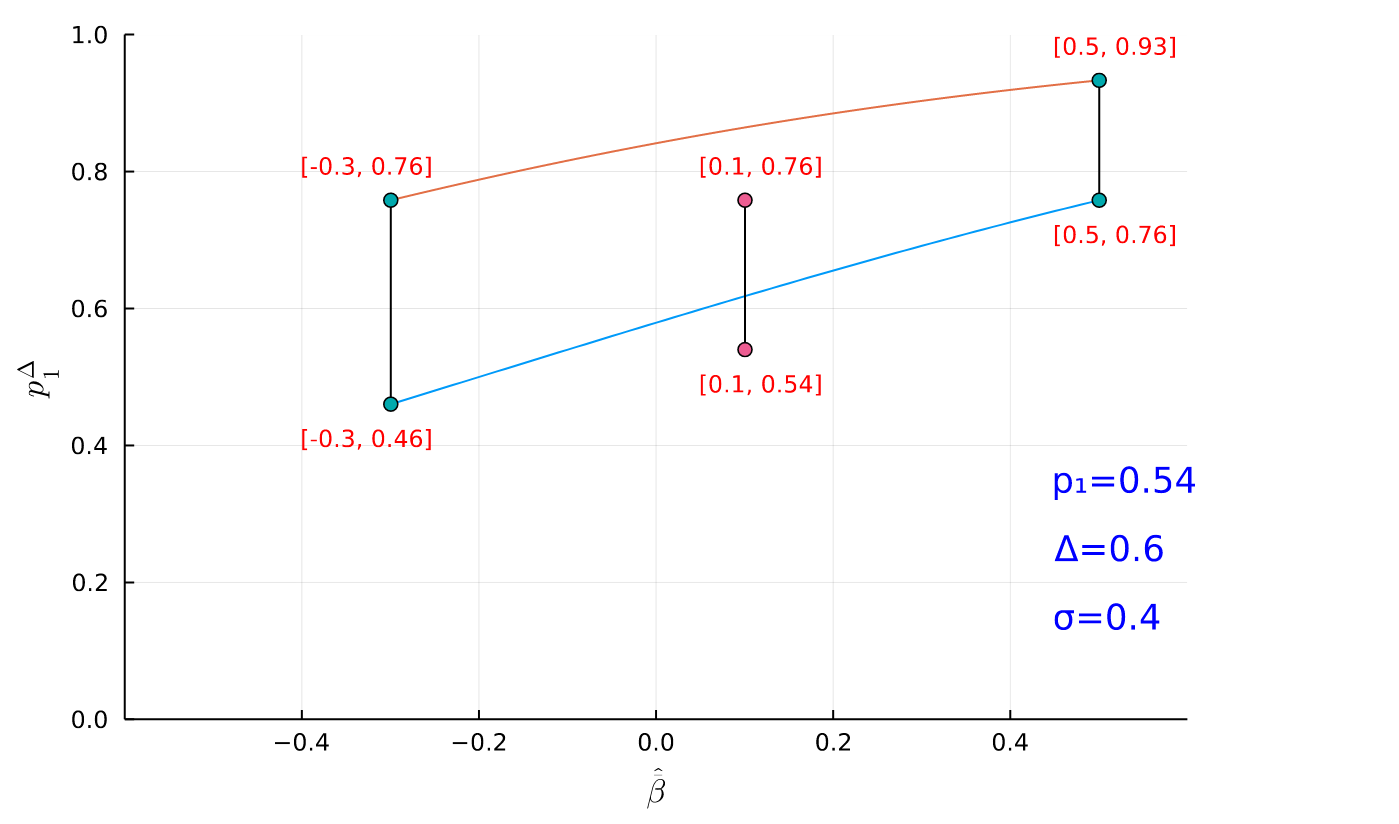}
\centering
\caption{Identification region for $p_1^{\Delta}$ when $2\sigma>\Delta$}
\end{figure}

\begin{figure}[ht]
\includegraphics[scale=0.53]{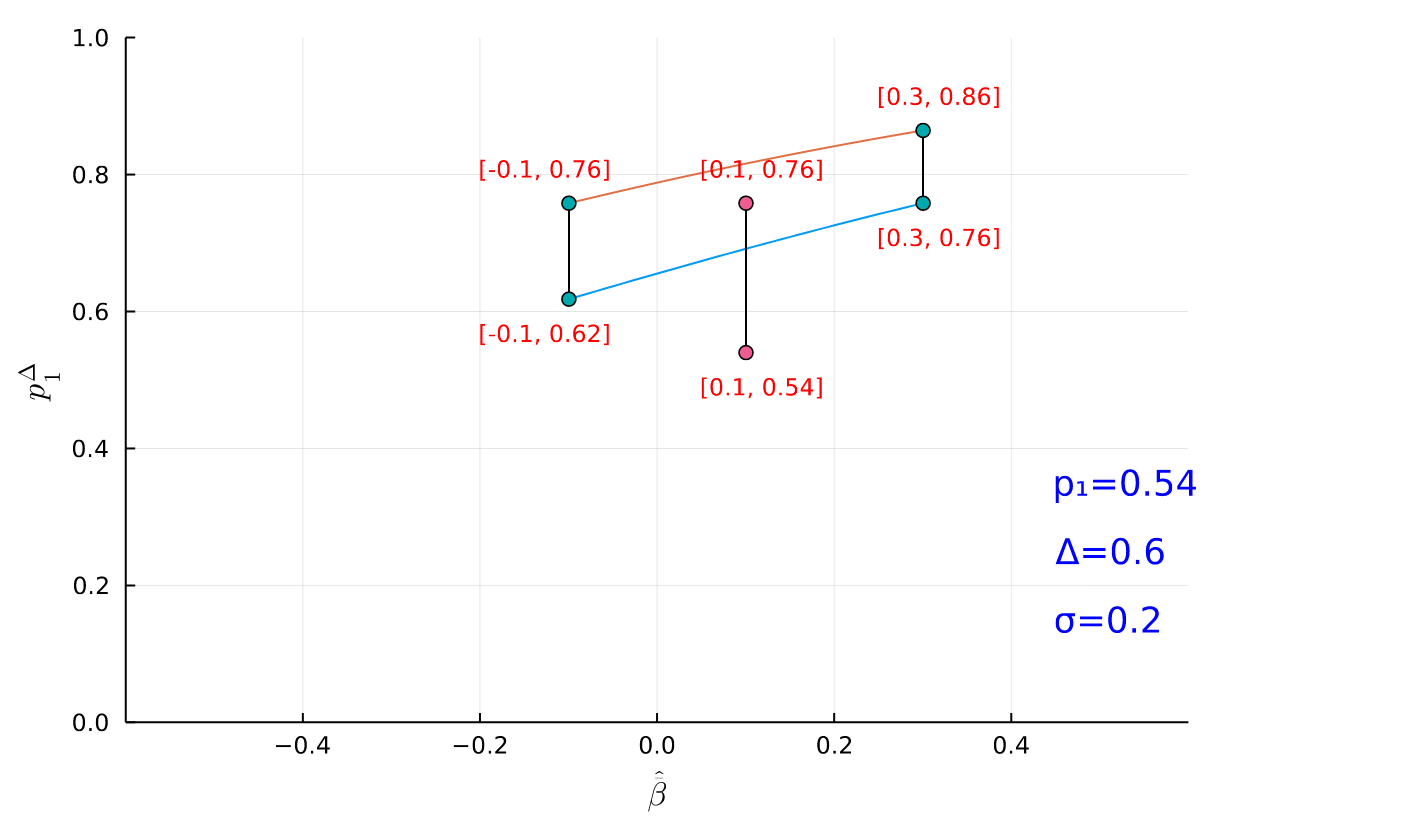}
\centering
\caption{Identification region for $p_1^{\Delta}$ when $2\sigma<\Delta$}
\end{figure}

\newpage

\bibliographystyle{ecta}
\bibliography{ambiguity}

\end{document}